\documentstyle[preprint,aps,prl,psfig]{revtex}
\tighten
\draft
\begin{document}

\draft

\title{The relaxation dynamics of a viscous silica melt: 
       II The intermediate scattering functions}

\author{J\"urgen Horbach and Walter Kob \footnote{Author to whom correspondence 
should be addressed. Permanent address: Laboratoire des Verres, 
Universit\'e Montpellier II, 34095
Montpellier, France. e-mail: kob@ldv.univ-montp2.fr}}
\address{Institut f\"ur Physik, Johannes Gutenberg-Universit\"at,
Staudinger Weg 7, D--55099 Mainz, Germany}
\maketitle

\vspace*{-10mm}

\begin{abstract}
We use molecular dynamics computer simulations to study the relaxation
dynamics of a viscous melt of silica. The coherent and incoherent
intermediate scattering functions, $F_d(q,t)$ and $F_s(q,t)$, show
a crossover from a nearly exponential decay at high temperatures
to a two-step relaxation at low temperatures. Close to the critical
temperature of mode-coupling theory (MCT) the correlators obey in the
$\alpha$-regime the time temperature superposition principle (TTSP)
and show a weak stretching. We determine the wave-vector dependence of the
stretching parameter and find that for $F_d(q,t)$ it shows oscillations
which are in phase with the static structure factor.  The temperature
dependence of the $\alpha$-relaxation times $\tau$ shows a crossover
from an Arrhenius law at low temperatures to a weaker $T$-dependence
at intermediate and high temperatures. At the latter temperatures the
$T$-dependence is described well by the power law proposed by MCT with
the same critical temperature that has previously been found for the
diffusion constant $D$ and the viscosity. We find that the exponent
$\gamma$ of the power law for $\tau$ are significantly larger than the
one for $D$. The wave-vector dependence of the $\alpha$-relaxation
times for $F_d(q,t)$ oscillates around $\tau(q)$ for $F_s(q,t)$ and
is in phase with the structure factor.  Due to the strong vibrational
component of the dynamics at short times the TTSP is not valid in the
$\beta$-relaxation regime. We show, however, that in this time window the
shape of the curves is independent of the correlator and is given by a
functional form proposed by MCT. We find that the value of the von
Schweidler exponent and the value of $\gamma$ for finite $q$ are
compatible with the expression proposed by MCT.  Finally we discuss
the $q$-dependence of the critical amplitude and the correction term and
find that they are qualitatively similar to the ones for simple liquids
and the prediction of MCT.  We conclude that, in the temperature regime
where the relaxation times are mesoscopic,  many aspects of the dynamics
of this strong glass former can be rationalized very well by MCT.

\end{abstract}


\section{Introduction}
\label{sec1}

If a glass forming liquid is cooled from high to low temperatures
one finds that its relaxation time $\tau$, or its viscosity $\eta$,
increases rapidly by many decades~\cite{books}. If $\eta(T)$ is
plotted in an Arrhenius plot, i.e. $\log(\eta)$ versus $1/T$, one
finds that the shape of the curves is not universal but depends on the
material. For two extreme cases of the shape, Angell has coined the terms
``strong'' and ``fragile'' glass formers~\cite{angell85}. Strong glass
formers are liquids whose $\eta(T)$ curve shows an Arrhenius law in
the whole accessible temperature range. In contrast to this, fragile
liquids show a pronounced crossover at intermediate temperatures from an
Arrhenius-like law at high temperatures to an Arrhenius-like law at low
temperatures with a {\it higher} activation energy. It is one of the
merits of the so-called mode-coupling theory (MCT) of the glass
transition~\cite{mct,gotze99} to offer an explanation for the existence of
this crossover in terms of non-linear dynamical feedback effects and to
make detailed predictions for the dynamics of the glass forming liquids
close to this crossover temperature $T_c$, the critical temperature
of MCT. In the past years there have been a large number of attempts
to check the validity of these predictions and the results of these
efforts was that the theory is indeed able to give a surprisingly good
description of the dynamics in this temperature range~\cite{gotze99}. It
has to be emphasized that some of these tests concerned not only the
predictions of the theory on a {\it qualitative} level, but also on a
{\it quantitative} one, and that also the outcome of these tests were often
in very good agreement with the theoretical prediction. Thus one can say
that the theory is able to describe in the vicinity of $T_c$ the dynamics
of fragile liquids on a qualitative as well as a quantitative level.

Things are much less clear for the case of {\it strong} glass formers since
by definition they do not show a crossover between two temperature
dependences in $\eta(T)$, or only a very mild one. Hence it is {\it a
priori} not clear at all whether these types of systems have a crossover
temperature $T_c$ and hence whether there is a temperature regime in
which MCT can be used to describe their dynamics. It was therefore a
bit surprising when it was found that glycerol, a glass former which
is a rather strong glass forming liquid, shows a dynamics which can be
described well by means of this theory~\cite{glycerol,franosch97}. Very
recently a computer simulation study of liquid SiO$_2$, the paradigm
of a strong glass former, showed that the dynamics of this system 
at high temperatures
shows features which are very reminiscent of the dynamics of fragile systems
around their $T_c$~\cite{horbach99}. In particular it was shown that
also in silica a critical temperature $T_c$ can be identified, with a
value around $T_c=3330$~K, i.e. a temperature which is presently outside
the reach of experimental investigations. It is of interest to note,
however, that for this material an extrapolation of experimental data
to higher temperatures gives evidence that around 3220~K a bend in the
viscosity would indeed be observed~\cite{hess96}, thus supporting the
results from the simulation. Very recently it has also been demonstrated
that MCT can be used to understand the dynamics of liquid silica not only
on a qualitative level, but also on a quantitative one. E.g. it was shown
that the theory is able to predict with good accuracy quantities like the
wave-vector dependence of the Debye-Waller factor~\cite{sciortino01}.
The only input in that calculation were static quantities, i.e.~the
partial structure factors and static three--point correlation functions,
which were determined directly from simulations.  
Also for water, another network forming liquid, it was found that MCT predicts
the wave-vector dependence of the Debye-Waller factor with a very good 
accuracy~\cite{fabbian99}.
Thus these studies give evidence that MCT is also able to give a correct 
description of the dynamics of strong liquids
in the temperature region where the dynamics starts to slow down significanly,
hence expanding the range of applicability of the theory considerably.

\mbox{From} the point of view of the theory this possibility is perhaps
not that surprising. When the theory was developed originally, some
terms in the equations of motions were dropped in order to simplify the
equations and because it was argued that they are not very relevant to
understand the main predictions of the theory~\cite{bengtzelius84}. This
``ideal version'' of the theory has subsequently been tested extensively
in experiments and computer simulations~\cite{mct,gotze99}. A bit later it
was recognized, however, that close to the critical temperature $T_c$
the terms that were dropped will change some of the predictions of
the theory considerably~\cite{das86,gotze87}, since they are needed
to describe processes, today called ``hopping processes'', that become
relevant at low temperatures. If these hopping processes are very small,
the predicted temperature dependence of the relaxation times shows a
strong bend in an Arrhenius plot, reminiscent to the one seen in fragile
glass formers~\cite{gotze87}. If these processes are very pronounced
this temperature dependence shows only a weak bending around $T_c$,
thus a similar behavior as the one found in strong glass formers. Thus
it seems that {\it a priori} the theory is indeed able to describe also strong
glass formers. 

However, it is presently not clear whether the extended form of
the theory, i.e.~the version of the theory which tries to include
hopping effects, is really reliable also beyond a phenomenological
description. Therefore it is important to check to what extent the
theoretical predictions survive in the case where one deals with a real
material or a realistic microscopic model in which hopping effects are
expected to be important. Such investigations might also give an idea
how the theory can be improved or extended in order to describe the
dynamics of realistic glassforming liquids. The present paper is hence
a contribution to this issue.  For this we use 
molecular dynamics computer simulations to investigate the dynamics of
liquid silica in great detail. In particular we will study to what extent
the behavior predicted by the theory can be found in this dynamics,
i.e. we test in a stringent way to what extent MCT is able to give a
reliable description of the relaxation dynamics of strong glass formers
at high temperatures.

The rest of the paper is organized as follows. In the next section we give
the details on the model we use and the simulations. The following section
is devoted to the presentation of the results, which are summarized and
discussed in the last section.

\section{Model and Details of the Simulation}
\label{sec2}
In this section we present the model we used to describe the interaction
of silica and give some of the details of the simulation. Further
information on the latter can be found in Ref.~\cite{horbach_diss}.

In order to obtain a realistic description of the structural and dynamical
properties of a given material by means of classical simulations, it
is necessary to have a potential at hand which describes the interactions
between the different ions realistically. Due to the importance of silica
in applied as well as fundamental science it is nowadays possible to
find a multitude of different potentials in the literature, most of
them having certain advantages and disadvantages. Various simulations
have shown that one of the most reliable of these potentials is the one
proposed by van Beest, Kramer, and van Santen (BKS) which was obtained by
determining the energy surface of a SiO$_4$H$_4$ tetrahedron by means of
{\it ab initio} calculations, parametrizing this surface with a simple
analytical expression and subsequently doing some lattice dynamics
calculations with it to improve the fit parameters~\cite{beest90}.

The functional form of the BKS potential is
\begin{equation}
\phi_{\alpha \beta}(r)=
\frac{q_{\alpha} q_{\beta} e^2}{r} +
A_{\alpha \beta} \exp\left(-B_{\alpha \beta}r\right) -
\frac{C_{\alpha \beta}}{r^6}\quad \alpha, \beta \in
[{\rm Si}, {\rm O}],
\label{eq1}
\end{equation}

\noindent
where $r$ is the distance between the ions of type $\alpha$ and $\beta$.
The values of the constants  $q_{\alpha}, q_{\beta}, A_{\alpha
\beta}, B_{\alpha \beta}$, and $C_{\alpha \beta}$ can be found in
Ref.~\cite{beest90}. For the sake of computational efficiency the short
range part of the potential was truncated and shifted at 5.5~\AA. This
truncation has also the benefit to improve the agreement between the
density of the amorphous glass at low temperatures as determined from
the simulation with the experimental value. Previous simulations
have shown that this potential is able to reproduce a variety of
properties of real amorphous silica, such as the density anomaly,
the static structure factor, the specific heat, the viscosity and the
thermal conductivity~\cite{bks_sim}.  Thus it can be expected that it is
also sufficiently accurate to give a reliable description of the time
dependence of the intermediate scattering function, the central quantity
discussed in the present paper.

The system investigated has 8016 ions in a cubic box with fixed size
$L=48.37$~\AA. This size is sufficiently large to avoid the finite
size effects found in the dynamics of such systems~\cite{horbach96,horbach01}.
The equations of motion have been integrated by means of the velocity
form of the Verlet algorithm, using a time step of 1.6~fs. The Coulombic
part of the potential has been evaluated by means of the Ewald sum using a
constant $\alpha L$=12.82. Before the microcanonical production runs were
started the system was equilibrated by coupling it to a stochastic heat
bath. The duration of this equilibration was significantly longer than
the typical relaxation times of the system. Therefore we are sure that
all the results presented below reflect the relaxation dynamics {\it in
equilibrium}, i.e. that they are not affected by aging phenomena. The
temperatures investigated were 6100~K, 5200~K, 4700~K, 4300~K, 4000~K,
3760~K, 3580~K, 3400~K, 3250~K, 3100~K, 3000~K, 2900~K, and 2750~K. In
order to improve the statistics of the results we averaged at each
temperature over two independent runs.

Note that all these temperatures are {\it above} the melting
temperature of real silica, which is around 2000~K~\cite{heaney}, a
temperature which the BKS potential is able to reproduce reasonably
well~\cite{roder}. Thus none of our simulations probe the dynamics of
the system in its supercooled state. Nevertheless we will see below that
even at these relatively high temperatures the dynamics of the system
is very slow. Thus this shows that in order to have a slow relaxation
dynamics it is not necessary to be in a supercooled state.

\section{Results}
\label{sec3}

In this section we will present the results. In the first part we will
discuss the relaxation dynamics of the system at long times, i.e.~the
$\alpha-$relaxation regime.  In the second part we will focus on
intermediate times, i.e.~the $\beta-$relaxation regime, and compare
these results with the predictions of MCT.

The quantity of main interest which will be studied in the present paper
is the intermediate scattering function $F(q,t)$ and its self part,
$F_s(q,t)$~\cite{hansenmcdonald86}. These two space-time correlation
functions can be defined most easily in terms of the density fluctuations
for the particles of type $\alpha$

\begin{equation}
\delta \rho_{\alpha}({\bf q},t) = \sum_{j=1}^{N_{\alpha}} \exp(i {\bf q} 
\cdot {\bf r}_j(t)) ,
\label{eq2}
\end{equation}

\noindent
where ${\bf q}$ is the wave-vector. The intermediate scattering function
is then defined as

\begin{equation}
F^{\alpha\beta}(q,t) = \frac{1}{N_{\rm Si}+N_{\rm O}} \langle \delta
\rho_{\alpha}({\bf q},t) \delta \rho_{\beta}^*({\bf q},0) \rangle \quad .
\label{eq3}
\end{equation}

\noindent
Here $\langle . \rangle$ stands for the thermal average and
we have assumed that the system is isotropic and hence $F(q,t)$
depends only on the modulus $q=|{\bf q}|$. From Eqs.~(\ref{eq2}) and
(\ref{eq3}) it is clear that $F(q,t)$ can be split into two parts,
$F^{\alpha\beta}(q,t)=F_s^{\alpha}(q,t)\delta_{\alpha\beta}
+F_d^{\alpha\beta}(q,t)$, where the functions $F_s^{\alpha}$ and
$F_d^{\alpha\beta}$ correspond to the diagonal and off diagonal elements
in the double sum of $F^{\alpha\beta}$:

\begin{equation}
F_s^{\alpha}(q,t) = \frac{1}{N_{\alpha}} \sum_{j=1}^{N_{\alpha}} 
\langle \exp( i {\bf q} \cdot 
({\bf r}_j(t)-{\bf r}_j(0))) 
\rangle
\label{eq4}
\end{equation}

\noindent
and 

\begin{equation}
F_d^{\alpha\beta}(q,t) = \frac{1}{N_{\rm Si}+N_{\rm O}}
\sum_{j=1}^{N_{\alpha}} 
\sum_{k=1}^{N_{\beta}}{}^{\prime}
\langle \exp( i {\bf q} \cdot ({\bf r}_j(t)) -{\bf r}_k(0))\rangle \ .
\label{eq5}
\end{equation}

\noindent
Here the prime means that the term with $k=j$ has to be left out.

Note that $F_s(q,t)$ and $F_d(q,t)$ are the space-Fourier
transforms of the self and distinct part of the van Hove correlation
function~\cite{hansenmcdonald86}, thus quantities which are often
studied in computer simulations of liquids. However, since in scattering
experiments the latter functions are not directly accessible, in contrast
to $F_s(q,t)$ and $F_d(q,t)$, and since also from a theoretical point
of view the scattering functions are more interesting than the van Hove
functions, it is appropriate to determine also them in simulations.
We emphasize that to obtain the results of the present work we have not
used the connection between the van Hove functions and the intermediate
scattering functions, but calculated the latters directly from their
definitions given in Eqs.~(\ref{eq4}) and (\ref{eq5}). For this we
averaged over all $\bf{q}$-vectors that had the same modulus and that
were compatible with the size of the simulation box.

\begin{figure}[h]
\psfig{file=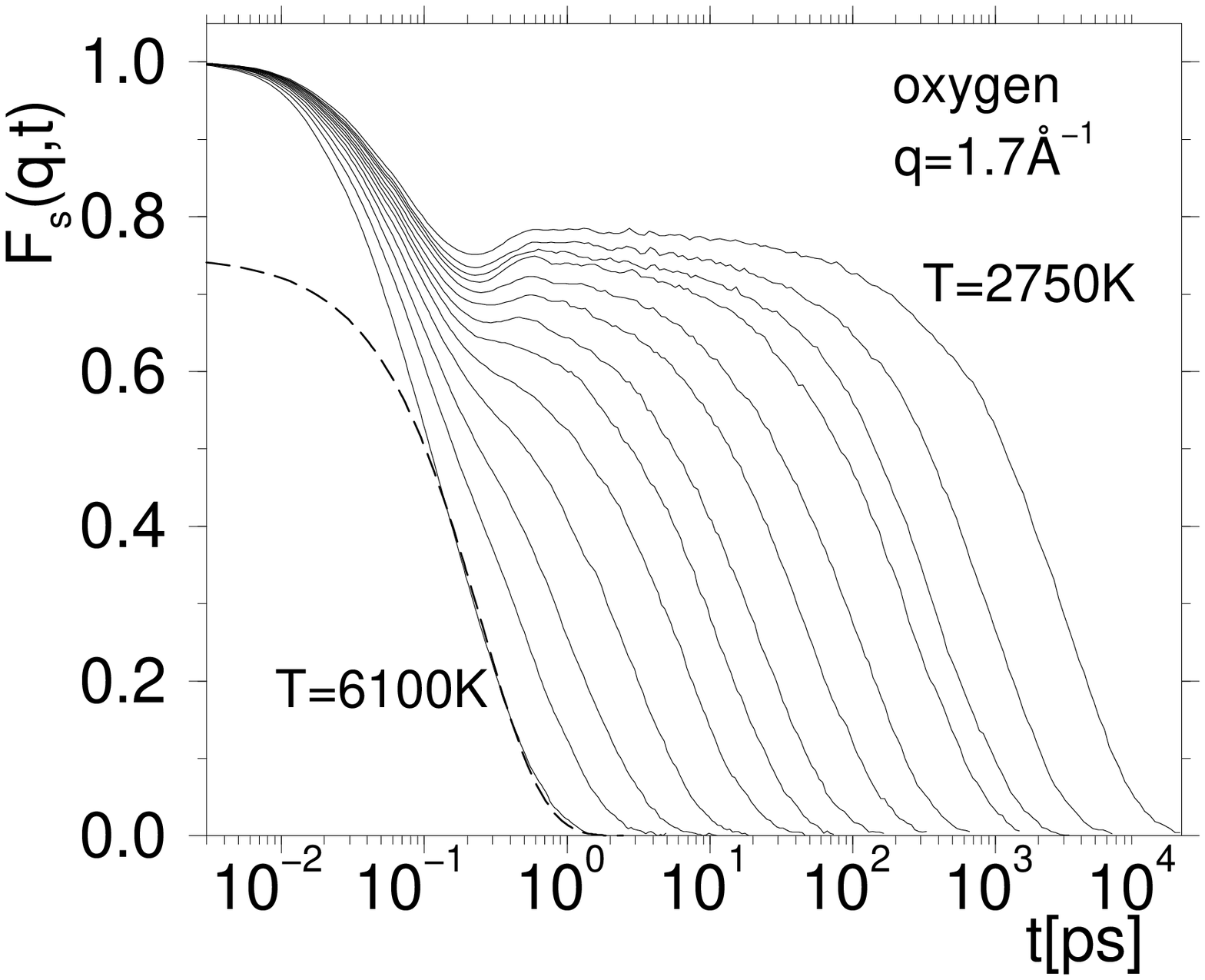,width=11cm,height=8cm}
\vspace*{2mm}

\caption{
Time dependence of the incoherent intermediate scattering function for the
oxygen atoms for all temperatures investigated.  The wave-vector $q$
is 1.7~\AA$^{-1}$, the location of the first peak in the structure factor.
The dashed line is a fit to the curve for 6100~K with a exponential function 
(see text for details). 
} 
\label{fig1}
\end{figure}

In Fig.~\ref{fig1} we show the time dependence of $F_s(q,t)$ for the
oxygen atoms for all temperatures investigated. The wave-vector is
$q=1.7$~\AA$^{-1}$, the location of the first sharp diffraction peak
in the structure factor, i.e. the length scale that corresponds to
the typical distance between two adjacent tetrahedra in the network~\cite{horbach99}.
(We mention that for other wave-vectors the curves look
qualitatively very similar.)  From this graph we see that at high
temperatures the curves show at very short times a crossover from a
ballistic regime to a relaxation behavior which is basically exponential
and that the correlation function decays to zero within one ps,
i.e. very rapidly.
To illustrate this exponential decay we have included in Fig.~\ref{fig1} 
an exponential function 
$f(t)=0.75 \exp(-t/0.25 {\rm ps})$, dashed line, which essentially coincides
with the simulation curve at $T=6100$~K for $t>0.1$~ps. (We emphazise
that due to the complex vibrational dynamics of the system at short
times~\cite{horbach01}, one has to use an amplitude significantly smaller
than 1.0)
For intermediate and low temperatures one observes
two additional regimes: Immediately after the ballistic regime,
which last just up to $\approx 0.02$~ps, the motion of the ions shows
a vibrational character, as can be inferred from the presence of a
dip in the correlator at around 0.2~ps. This type of dynamics is the
dominant one for temperatures below the glass transition, but from the
figure we recognize that it can be seen even at temperatures as high as
3580~K. From a microscopic point of view this dynamics corresponds to
the rattling motion of the ions in the cage formed by their surrounding
neighbors and the network. (Note that this cage is not rigid at all,
since the particles that form the ``walls'' of this cage move themselves
also.) Although every liquid will show some sort of vibrational dynamics,
the one for SiO$_2$ is special in that it has also important contributions
at quite low frequencies (1-2~THz) the origin of which has in recent
years been the matter of a strong debate~\cite{horbach01,boson_peak}.
For times which are somewhat longer than the ones for the vibrations,
i.e.  $t\geq 0.2$~ps, this type of motion is damped out and the particles
slowly start to leave the mentioned cage. This dynamics, in the following
called $\beta$-process~\cite{mct,gotze99}, is thus the beginning of the
$\alpha$-process, i.e. the time regime in which the ions leave the cage
completely and lead to the structural relaxation of the system. In the
correlation functions it is seen as a plateau at intermediate times,
the length of which rapidly increases with decreasing temperature thus
pushing the $\alpha$-relaxation to large times (see Fig.~\ref{fig1}). MCT
makes detailed predictions on the dynamics of the system on the time
scale of the $\beta$-relaxation and in the following we will check the
validity of these in detail.

\begin{figure}[h]
\psfig{file=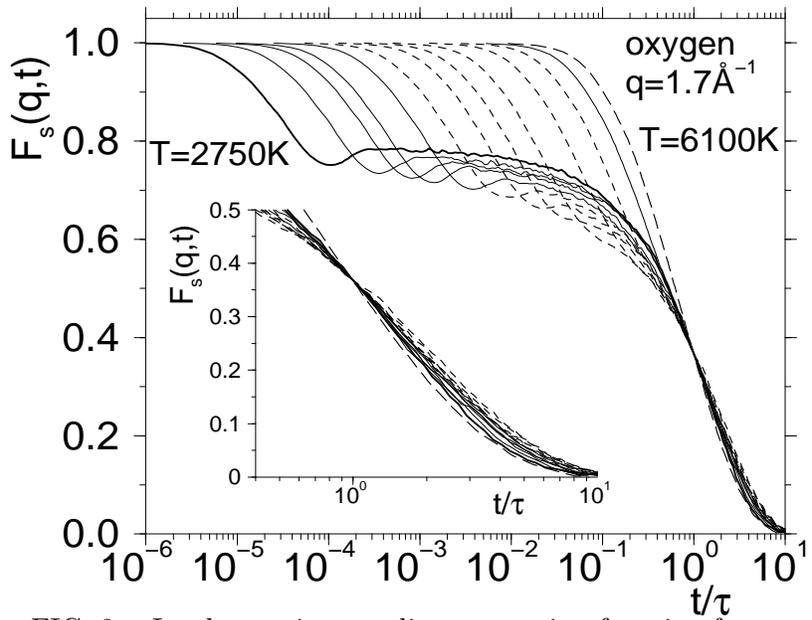,width=11cm,height=8cm}
\caption{
Incoherent intermediate scattering function for oxygen versus $t/\tau(T)$,
where $\tau$ is the $\alpha$-relaxation time at temperature $T$. Inset:
Enlargement of the curves at long rescaled time.
See text for the meaning of the different line styles.
}
\label{fig2}
\end{figure}

\mbox{From} Fig.~\ref{fig1} one gets the impression that the shape of the
curves {\it in the $\alpha$-relaxation regime} does not depend on
temperature, i.e.~that the correlators obey the so-called time temperature
superposition principle (TTSP). This means that a correlator $\phi(t,T)$
can be expressed by

\begin{equation}
\phi(t,T)= \hat{\phi}(t/\tau(T)),
\label{eq6}
\end{equation}

\noindent
where $\tau(T)$ is the typical time scale for the decay of the correlation
function, i.e.~the $\alpha$-relaxation time, and $\hat{\phi}$ is a master
function. MCT predicts that there exists a so-called critical temperature
$T_c$ in the vicinity of which the TTSP is valid. Since this is one of
the main predictions of the theory it is of interest to check whether
or not it is valid for the present system. For this we have defined
an $\alpha$-relaxation time $\tau$ by requiring that at time $\tau$
the correlator has decayed to $e^{-1}$ of its initial value. If the
TTSP holds, a plot of the correlators at different temperatures versus
$t/\tau(T)$ should give a master curve. In Fig.~\ref{fig2} we show this
type of plot and we recognize that apparently the scaling does not work very
satisfactory. To show this we have plotted in the inset of the figure
an enlargement of the main figure at long rescaled times. From this
inset we recognize that the fact that the TTSP is not fulfilled is
not due to the noise in the data but rather a systematic trend in the
curves. We have also made sure that this violation of the TTSP is not
due to the way we have defined $\tau$, since a different definition gave
qualitatively the same results~\cite{horbach_diss}. Recall, however,
that the TTSP is predicted to hold only
slightly above $T_c$, which for the present system is around 3330~K~\cite{horbach99}. 
In fact, if the curves for the different temperatures are inspected 
carefully one finds that the ones in the temperature range 
$4700K\ge T \ge 3400K$ (short dashed lines in Fig.~\ref{fig2}) do 
indeed fall onto a master curve whereas the ones for high 
($T=6100$~K) and low ($T \leq 3000$~K) temperatures decay faster. Hence
the curve at $T=2750$~K (bold line in Fig.~\ref{fig2}) is already
very close to the exponentially decaying one at $T=6100$~K (long dashed
line in Fig.~\ref{fig2}). Hence we see that slightly above $T_c$
the correlation functions do indeed obey the TTSP, whereas it is 
violated if the temperature is too far below $T_c$.
This behavior is in qualitative agreement with the theory, since it is
predicted that far above and far below $T_c$ the relaxation is of Debye
type~\cite{mct,gotze87}.

We also mention that according to MCT the TTSP is supposed to hold also
in the $\beta$-relaxation regime, i.e.~in the time window in which
the correlation functions are close to the plateau. From the graph
we see, however, that in this time regime the curves do not collapse
at all. The reason for this is likely the fact that in this system the
microscopic dynamics, i.e.~the above mentioned low-frequency vibrations,
are so pronounced that they completely obscure the TTSP in this time
window~\cite{franosch98}, in contrast to simple liquids whose
vibrational dynamics does not have strong contributions at very low
frequencies, i.e.~around 1-2~THz.
Below we will see, however, that certain predictions of
MCT concerning the dynamics in the $\beta$-relaxation regime are still
valid, despite the presence of the strong vibrational dynamics.

\begin{figure}[h]
\psfig{file=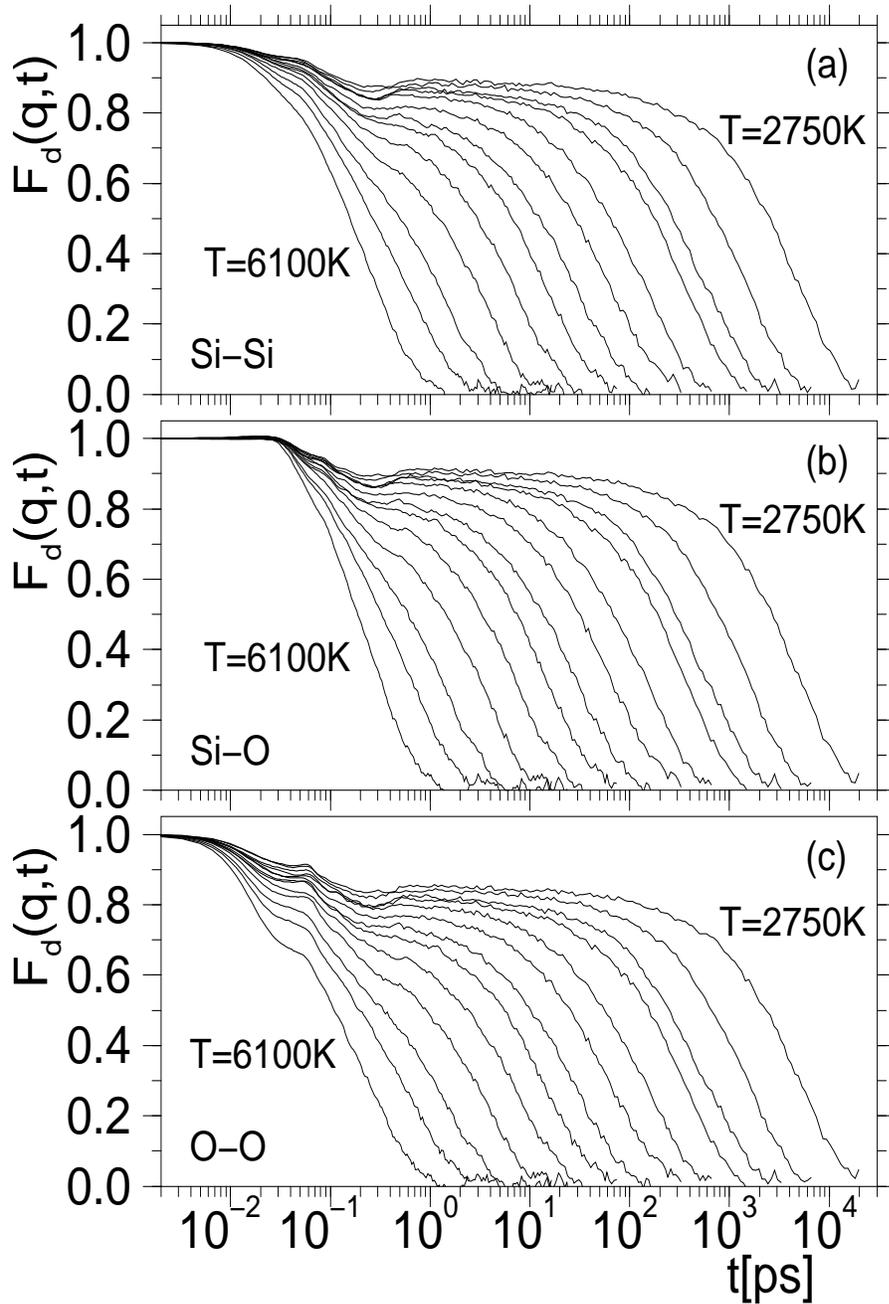,width=12cm,height=17cm}
\vspace*{2mm}

\caption{
Time dependence of the coherent intermediate scattering function for all
temperatures investigated. The wave-vector is 1.7~\AA$^{-1}$, the location of
the first peak in the structure factor. a) Si-Si, b) Si-O, c) O-O.
}
\label{fig3}
\end{figure}

The results presented so far concerned the incoherent part of the
intermediate scattering function of the oxygen atoms. For the case of
the silicon atoms the time and temperature dependence is qualitatively
similar, although for this species the TTSP at long times is fulfilled
a bit better~\cite{horbach_diss}.  Since $F_s(q,t)$ measures essentially
the dynamics of a tagged particle within its cage and how it escapes
this cage at long times, not much can be learned from this correlation
function on the {\it relative} motion of the particles. Therefore it is of
interest to study also the coherent intermediate scattering function
$F_d(q,t)$ which is defined in Eq.~(\ref{eq5}). In Fig.~\ref{fig3} we show
the time dependence of the three partial $F_d^{\alpha\beta}(q,t)$ for
all temperatures investigated.  The wave-vector is again 1.7~\AA$^{-1}$,
i.e. the location of the first sharp diffraction peak. Although from
a qualitative point of view these correlation functions are similar
to the incoherent ones, a closer inspection does show interesting
differences. E.g.~the coherent curves show the effects of the complex
vibrational dynamics of the system much more pronounced in that, e.g.,
several maxima and minima can be seen between 0.02~ps and 1~ps, whereas the
incoherent curves show only one local minimum (see Fig.~\ref{fig1}). We
also mention that the modulus of the incoherent functions as well as
$F_d^{\rm SiSi}$ and $F_d^{\rm OO}$ are always smaller than 1.0 since
they are (normalized) autocorrelation functions.  This is in contrast to
$F_d^{\rm SiO}$ which, being not a autocorrelation function, can become
larger than 1.0. A closer inspection of the curves for this correlator
does indeed show that for times around 0.03~ps the curves are larger than
1.0, which is the reason why this correlator is very flat at short times.

\begin{figure}[h]
\psfig{file=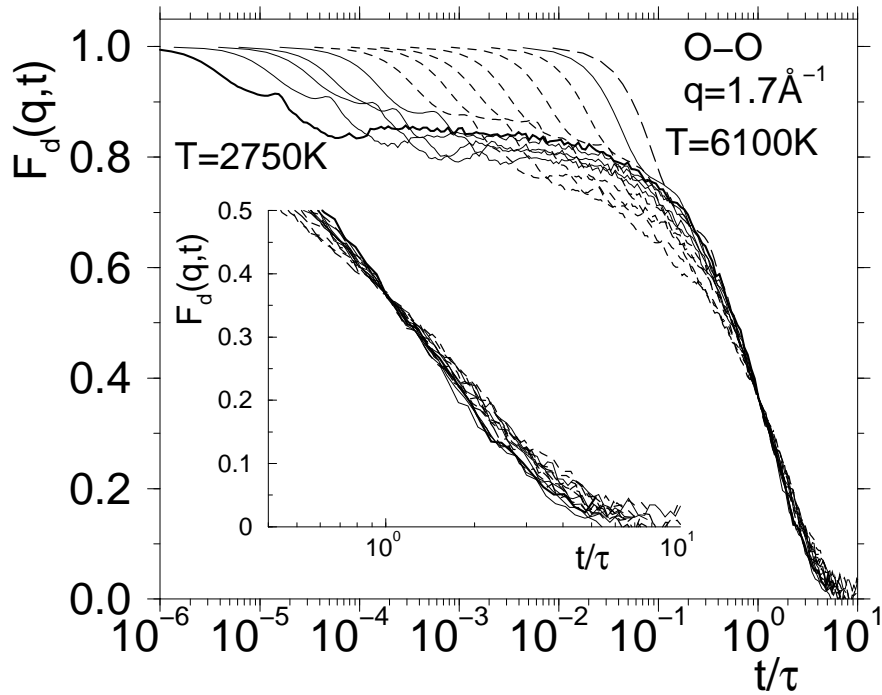,width=12cm,height=9cm}
\vspace*{2mm}

\caption{
Coherent intermediate scattering function for the oxygen-oxygen
correlation versus $t/\tau(T)$, where $\tau$ is the $\alpha$-relaxation
time at temperature $T$. Inset: Enlargement of the curves at long
rescaled time. See Fig.~\protect\ref{fig2} for meaning of the dashed curves.
}
\label{fig4}
\end{figure}

\mbox{From} Fig.~\ref{fig3} one sees that in the $\alpha$-relaxation regime
the shape of the curves depends only weakly on temperature, i.e.~that
the TTSP is valid. That this is essentially the case is demonstrated in
Fig.~\ref{fig4}, where we plot the coherent correlation functions for
the case of O-O versus the rescaled time $t/\tau$. We see that within
the noise of the data the curves basically collapse onto one master
curve. However, also in this case small systematic deviations from the
TTSP, as discussed for the incoherent functions, can be found upon
closer inspection of the curves. Thus this shows that also for these
types of correlation functions the TTSP holds only for temperatures
relatively close to $T_c$.

The intermediate scattering functions discussed so far can be
directly measured in spin echo experiments or photon correlation
experiments. However, in neutron or dynamic light scattering experiments
only the time-Fourier transforms of these correlation functions are
accessible. Therefore it is of interest to calculate these functions
also, in order to see to what extent the various regimes seen in the time
domain can be found in the frequency domain. Since the intermediate
scattering functions have been measured over more than seven decades
in time, the calculation of their Fourier transform is not a trivial
matter. In order to do this we have approximated each curve with a
spline under tension and calculated the Fourier transform of the spline
by means of the Filon formula~\cite{abramowitz}. Thus we obtained the
dynamic scattering functions $S(q,\omega)$ and $S_s(q,\nu)$. Since in the
experiments one often measures the imaginary part of the susceptibility,
$\chi''(q,\omega)$, we have multiplied these functions with $\omega$
to obtain $\chi''(q,\omega)$ and $\chi_s''(q,\omega)$, respectively.

\begin{figure}[h]
\psfig{file=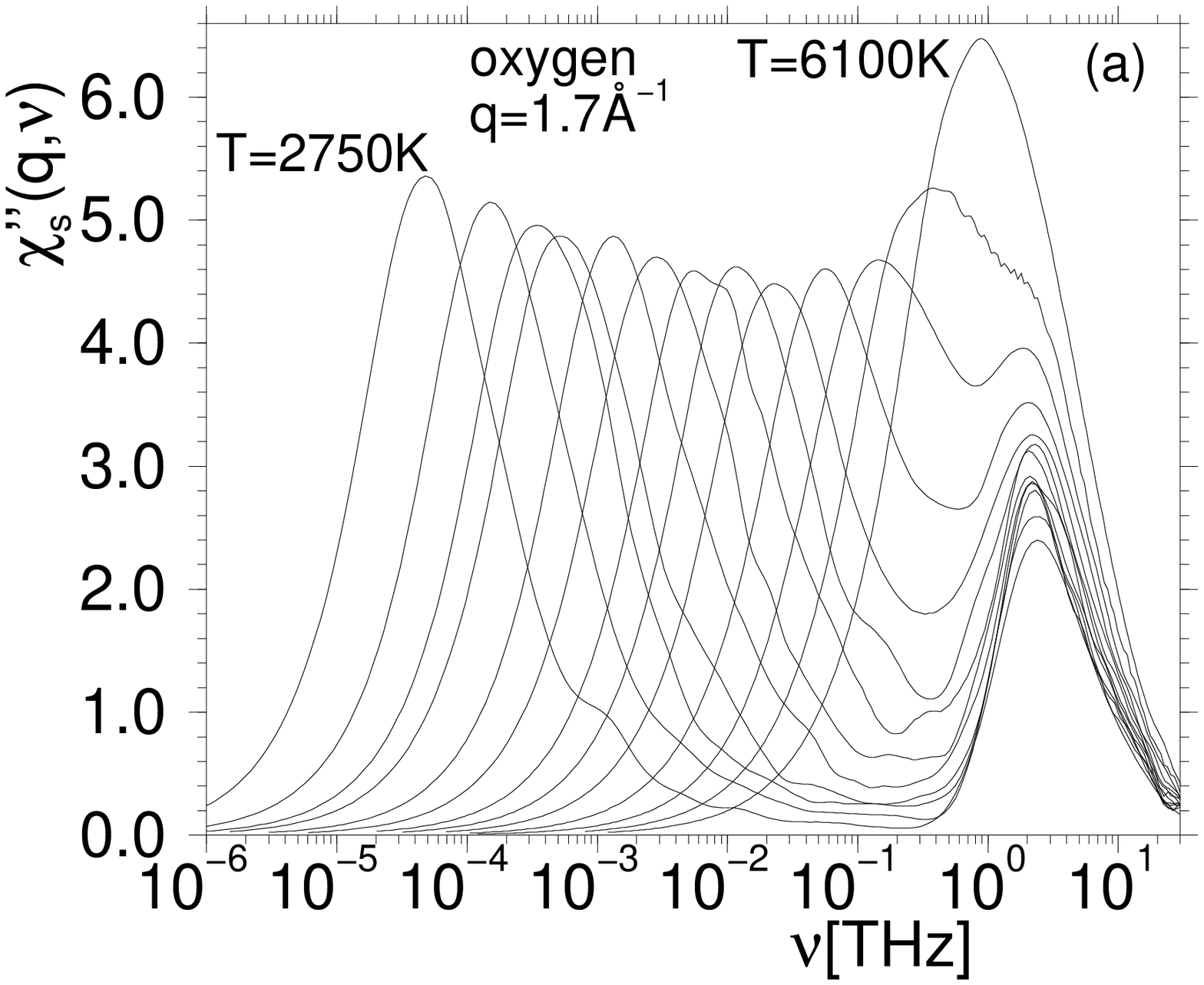,width=11cm,height=8cm}
\psfig{file=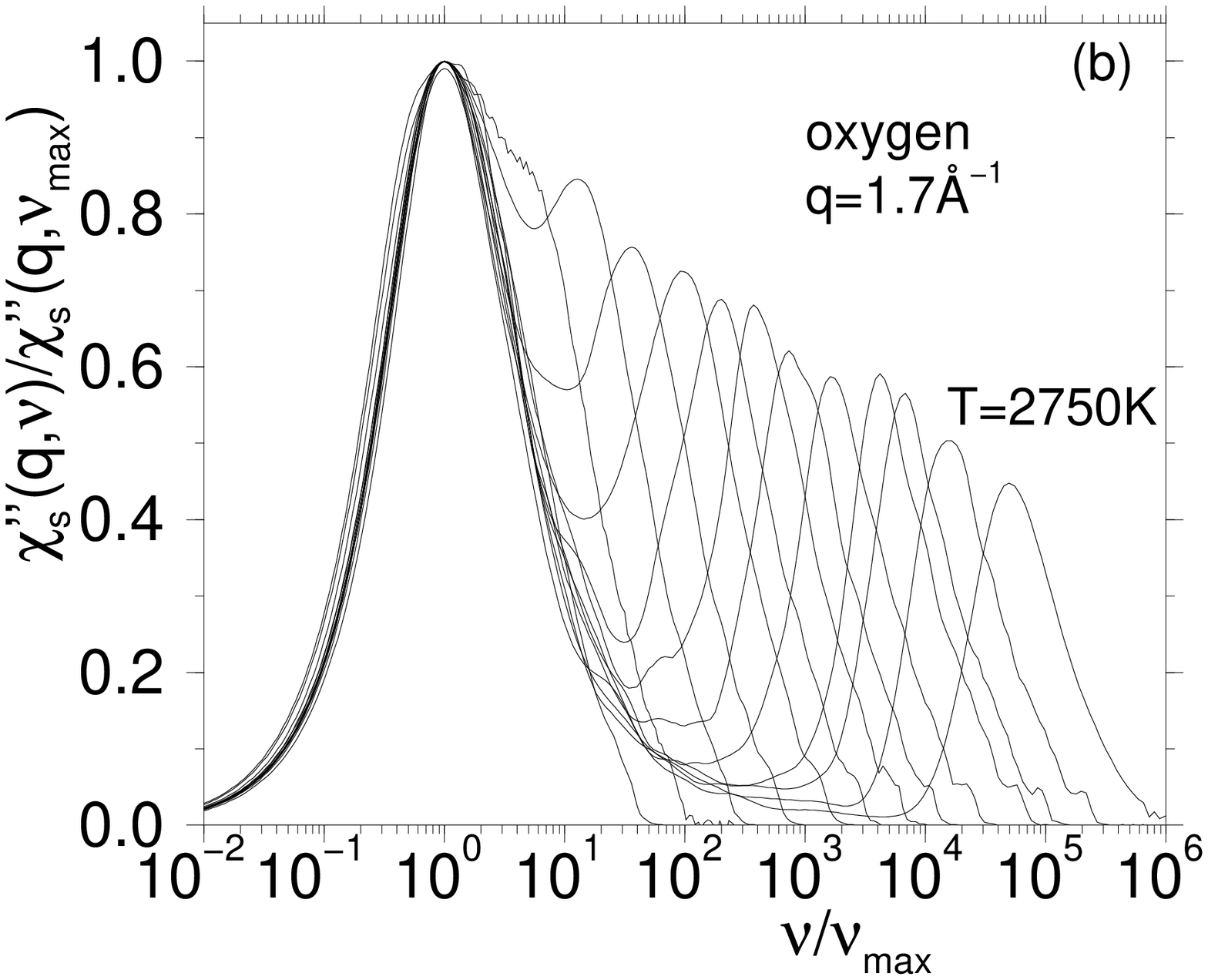,width=11cm,height=8cm}
\vspace{1mm}

\caption{
Frequency dependence of the imaginary part of the dynamic susceptibility
$\chi_s''(q,\nu)$ for the oxygen atoms for all temperatures
investigated.  a) $\chi_s''(q,\nu)$ versus frequency $\nu$. b)
$\chi_s''(q,\nu)/\chi_s''(q,\nu_{\rm max})$ versus $\nu/\nu_{\rm max}$,
where $\nu_{\rm max}$ is the location of the $\alpha$-peak in $\chi_s''$.
}
\label{fig5}
\end{figure}

In Fig.~\ref{fig5}a we show $\chi_s''$ as a function of $\nu=2\pi\omega$
for all temperatures investigated.  We see that at 6100~K $\chi_s''$
shows only a single broad peak. 
The width of this peak at half maximum is approximately 1.5 decades 
which shows that it is not Debye-like, since this would correspond
to a width of 1.14 decades. This increase is simply due to the fact that
the broad peak at $T=6100$~K is a sum of a microscopic and a relaxation
peak, where the relaxation peak should be of Debye type in agreement
with the exponential decay of the corresponding curve for $F_s(q,t)$
in Fig.~\ref{fig1} for $t>0.1$~ps.

If the temperature is lowered the single peak splits up into two, i.e.~into
a microscopic peak at high frequencies and an $\alpha$-peak at lower
frequencies. With decreasing temperature the amplitude of the microscopic peak
decreases but its location is independent of $T$.  In contrast to
this the $\alpha$-peak quickly moves to lower frequencies and also
its height increases slightly (to compensate the decrease in height
of the microscopic peak). However, the shape of the peak seems to be
essentially independent of $T$. At low temperatures we find a well defined
plateau between the $\alpha$-peak and the microscopic peak. A plot of
log($\chi_s''$) versus log($\nu$) shows that at these low temperatures
the whole spectrum can be described very well by a sum of a peak at
microscopic frequencies and an $\alpha$-peak with a high frequency wing
that scales like $\nu^{-y}$, where the exponent $y$ is $\approx 0.6$
and independent of $T$.  Thus this result is in agreement with the von
Schweidler law discussed below in the context of Fig.~\ref{fig12} which
gives an exponent 0.62. It is remarkable that this type of plot shows
that for temperatures around 3250~K, i.e.~around $T_c$, the susceptibility
seems {\it not} to be just the sum of an $\alpha$-peak and a microscopic peak,
but that near the minimum between the two peaks $\chi_s''(\nu)$ is
enhanced. This means that there is an additional process in this frequency
regime, the $\beta$-process of MCT. We emphasize that this enhancement is
only seen close to $T_c$, in agreement with the prediction of the theory.

That the shape of the $\alpha$-peak does indeed not depend on
temperature, is demonstrated in Fig.~\ref{fig5}b. Here we show
$\chi_s''(q,\nu)/\chi_s''(q,\nu_{\rm max})$ as a function of
$\nu/\nu_{\rm max}$, where $\nu_{\rm max}$ is the location of the
maximum of the $\alpha$-peak. From this plot we clearly see that within
the accuracy of our data the curves for low temperatures fall on top of
each other. Finally we mention that qualitatively the same results have
been obtained for the case of the silicon atoms.

\begin{figure}[h]
\psfig{file=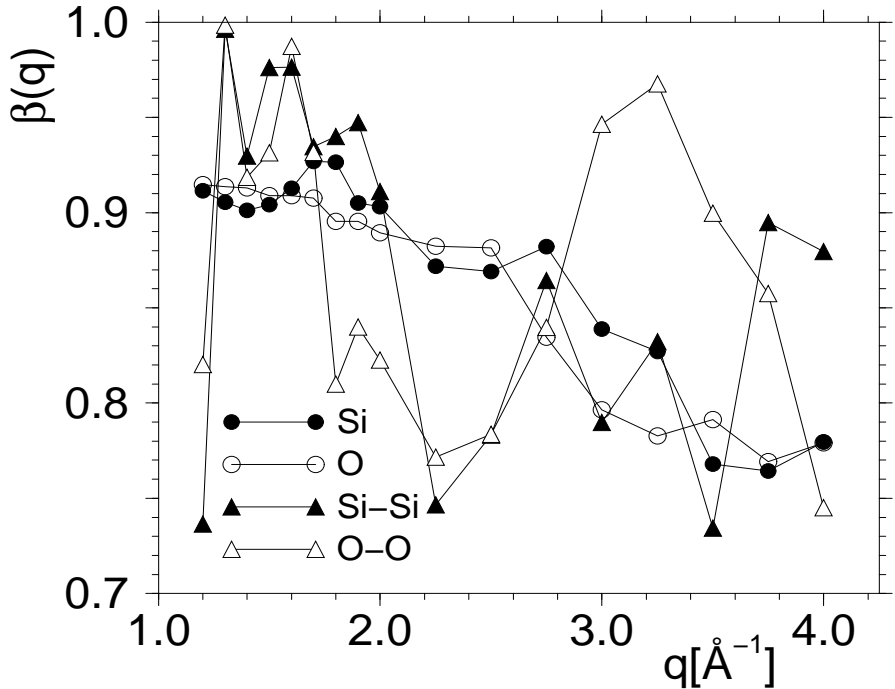,width=12cm,height=9cm}
\vspace{2mm}

\caption{
Wave-vector dependence of the Kohlrausch-Williams-Watts stretching parameter
for the silicon and oxygen incoherent intermediate scattering functions and
the silicon-silicon and oxygen-oxygen coherent intermediate scattering
functions. $T=2750$~K.
}
\label{fig6}
\end{figure}

Since the TTSP is valid to a high degree, we can determine the shape of
the master curves as a function of the observable. We have done this for
the case of the coherent and incoherent intermediate scattering function
for several values of $q$. To determine this shape we proceeded as
follows: For each correlator $\phi(t)$ we determined an $\alpha$-relaxation
time $\tau'$ at 2750~K by requiring that $\phi(t)=0.1$~\cite{footnote1}. We then fitted
$\phi(t)$ with a Kohlrausch-Williams-Watts (KWW) function 

\begin{equation}
\phi(t)=A \exp(-(t/\tau')^{\beta})
\label{eq7}
\end{equation}

\noindent
where the amplitude $A$ and the stretching parameter $\beta$ are free
fit parameters. This fit was done in the interval $0.02 \leq t/\tau'
\leq 2$ which for typical values of $\beta$ corresponds to about 90\%
of the height of the $\alpha$-relaxation (see Fig.~\ref{fig2}). We made
sure that choosing a different interval does not affect the results
significantly and we estimate the systematic error bar of $\beta$ due
to this choice to be usually smaller than 0.03~\cite{horbach_diss}. 
In Fig.~\ref{fig6} we show the wave-vector dependence of $\beta$ for the
two incoherent functions as well as for two coherent ones. We see that
for small and intermediate $q$ the value of $\beta$ is relatively large,
in agreement with the observation that strong glass-formers do not show
much stretching~\cite{boehmer}. However, for the large values of $q$, $\beta$
is significantly smaller than 1.0 since the curves for $F_s$ decrease
progressively with increasing $q$. This trend can be rationalized with the
prediction of MCT~\cite{fuchs93} that with increasing $q$ the value of
$\beta$ should approach the value of the von Schweidler exponent $b$,
discussed in more detail below, which for our system is around 0.62,
i.e. significantly smaller than 1.0.  At $q=4$~\AA$^{-1}$ the value
for $\beta(q)$ as determined from $F_s$ for silicon and oxygen is
around 0.77, i.e.~higher than the value predicted by the theory for
large $q$. However, it has to be realized that $q=4$~\AA$^{-1}$ cannot be
considered yet as ``large'' since at this wave-vector the static structure
factor shows still significant oscillations~\cite{horbach99}. In a MCT
calculation for a hard sphere system it has been found that only at
very large wave-vectors the value of $\beta$ becomes close to the von
Schweidler exponent, i.e. at $q$ at which the static structure factor
is essentially 1.0~\cite{fuchs92}. In view of this result it is hence not
surprising that the value of our $\beta$ at the largest $q$ is still
significantly above $b$.

The curves for the coherent functions oscillate around the ones for the
incoherent ones. This oscillation is in phase with the static structure
factor with peaks that correspond to the maxima in $S_{\alpha\alpha}(q)$.
Note that for large $q$ the coherent functions are approximated well by
the corresponding incoherent ones~\cite{hansenmcdonald86} and thus it
is expected that the values of $\beta$ become identical also. However,
in the range of $q$ for which we can reliably determine the value of
$\beta$ this is not yet the case.  This supports our argument that none
of the $\beta(q)$ curves has yet converged to its asymptotic value for
$q\to \infty$.  Note that qualitatively the same behavior for $\beta(q)$
has also recently been observed for water, where it was found that
$\beta(q)$ converges to the von Schweidler exponent $b$ for large $q$
and that it shows oscillation in phase with the static structure for
its coherent part~\cite{water}.

\begin{figure}[h]
\psfig{file=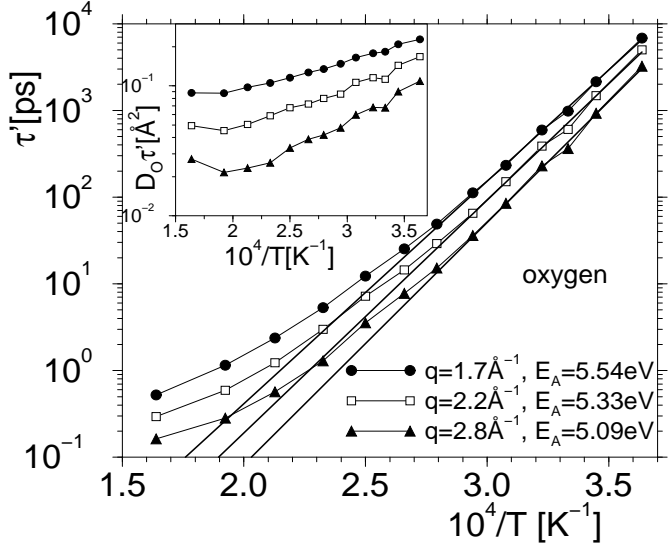,width=9cm,height=7cm}
\vspace{2mm}

\caption{
Temperature dependence of the $\alpha$-relaxation time $\tau'$ as determined
from $F_s(q,t)$ for oxygen for various values of $q$. The solid straight
lines are fits to the data at low temperature with an Arrhenius law
with the activation energies given in the figure. Inset: Temperature
dependence of the product $D_{\rm O}\tau'$, where $D_{\rm O}$ is the
diffusion constant for the oxygen atoms.
}
\label{fig7}
\end{figure}

We now address the temperature dependence of the $\alpha$-relaxation
time $\tau$. Since SiO$_2$ is a strong glass-former~\cite{angell85} one
expects that $\tau$ follows an Arrhenius law. In Ref.~\cite{horbach99}
we have shown, however, that for the diffusion constant as well as the
viscosity this law is found only at relatively low temperatures. For
intermediate and high temperatures a significant deviation was found
in that the temperature dependence was weaker than the one expected
from an Arrhenius law. That this is the case for $\tau(T)$ as well is
demonstrated in Fig.~\ref{fig7} where we show the $\alpha$-relaxation
time for $F_s(q,t)$ for the oxygen atoms (the curves for the silicon atoms
look qualitatively the same). The three curves that are shown correspond to
$q$-values at the first sharp diffraction peak ($q=1.7$~\AA$^{-1}$), the
location of the first minimum in the structure factor ($q=2.2$~\AA$^{-1}$),
and the location of the main peak in $S(q)$ ($q=2.8$~\AA$^{-1}$). We see
that, apart from a $q$-dependent prefactor, the temperature dependence
of the three relaxation times is very similar in that we find at low $T$
an Arrhenius law and at intermediate and high temperatures a crossover
to a weaker temperature dependence. The activation energies found at
low $T$ are between 5-5.5~eV, which compares well with the experimental
values for the diffusion constant and the viscosity (4.7~eV and 5.33~eV,
respectively~\cite{urbain82,mikkelsen84}). Thus we conclude that also
with regard to this quantity the BKS model is quite reliable.

In Ref.~\cite{horbach99} we showed that the deviations from the Arrhenius
law at intermediate and high temperatures are related to a change of
the transport mechanism of the ions in that at low $T$ the atoms make
independent hops whereas for high $T$ the motion is much smoother and
flow-like. Such a change in mechanism is predicted by MCT around the
critical temperature $T_c$~\cite{mct,gotze99}, which for our system has been
estimated to be around 3330~K~\cite{horbach99}. The theory predicts that
for temperatures a bit above $T_c$ the typical relaxation times increase
like a power-law

\begin{equation}
\tau(T) \propto (T-T_c)^{-\gamma}
\label{eq8}
\end{equation}

\noindent
with an exponent $\gamma$ that is universal for the system, i.e.~is
independent of the observable (Si or O, value of $q$, etc.). Thus the
bending of the curves in Fig.~\ref{fig7} can be rationalized by the
crossover from the power-law behavior given by Eq.~(\ref{eq8}) to an
Arrhenius behavior at low temperatures.

MCT predicts that the exponent $\gamma$ is independent of the observable
and that the diffusion constant $D$ should be proportional to $1/\tau$. In
order to check the validity of this prediction we have calculated the
product $D_{\rm O} \tau$ and plot it in the inset of Fig.~\ref{fig7}
as a function of $1/T$ for the three relaxation times shown in the
main figure. We see that although the product is basically constant
at high temperatures, it increases continuously at intermediate and
low temperatures. Since in this log-lin plot the curves at low $T$
become a straight line we conclude that the activation energies for
the diffusion constant and the relaxation times are not the same, a
conclusion that is confirmed if one measures these energies directly
(see below). Thus we conclude that the proportionality $D \propto
\tau^{-1}$ does not hold around $T_c$ or lower temperatures. One
reason for $D \not\propto \tau^{-1}$ might be the presence of dynamical
heterogeneities~\cite{dyn_het}, i.e.~that the cage that each particles
sees changes significantly from particle to particle which has the
effect that also the time to escape this cage depends strongly on
the particle. It is then easy to see that in this case the (average)
diffusion constant is not inversely proportional to the (average)
relaxation time~\cite{hodgdon93}. The observation that $D$ is not strictly
proportional to $\tau^{-1}$ has in the past already been made in other
systems for which the predictions of MCT hold extremely well not only on
a qualitative but also on a quantitative level~\cite{lj_kob}. Thus, 
since the prediction
of MCT that $D \propto \tau$ is only the result of a leading order calculation
of the theory, the fact that the product $D \tau$ is not completely constant
should not be taken as strong evidence that for the present system
the theory is not applicable.

\begin{figure}[h]
\psfig{file=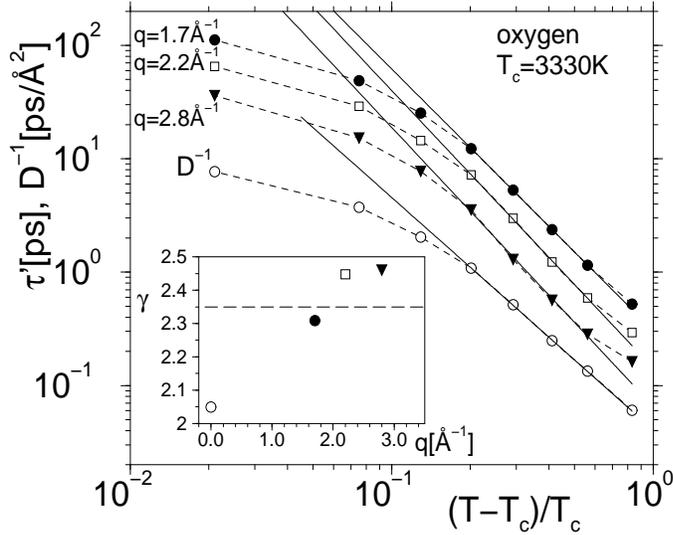,width=9cm,height=7cm}
\vspace*{2mm}

\caption{
The $\alpha$-relaxation time $\tau'$, as determined from $F_s(q,t)$ for
oxygen, as a function of $(T-T_c)/T_c$, where $T_c=3330$~K is the critical
temperature from MCT as determined in Ref.~\protect\cite{horbach99}. Also
included is the inverse of the diffusion constant as determined in
Ref.~\protect\cite{horbach99}. The solid straight lines are fits with
power laws of the form given by Eq.~(\protect\ref{eq8}). Inset: Value of
the exponent $\gamma$. The point at $q=0$ corresponds to the
exponent for the diffusion constant. The horizontal dashed line is the value
of $\gamma$ as determined from the dynamics in the $\beta$-relaxation regime.
}
\label{fig8}
\end{figure}

Since for the present system the temperature dependence of the
diffusion constant is not described so well by MCT, we investigate now the
$T$-dependence of the relaxation times directly. In Ref.~\cite{horbach99}
we determined the value of $T_c$ to be around 3330~K. Therefore we plot in
Fig.~\ref{fig8} the $\alpha$-relaxation time as a function of $T-T_c$,
for those temperatures which are higher than $T_c$. From this plot we
see that the data at high temperatures can indeed be described well with
a power-law (bold straight lines), in agreement with the prediction of the
theory (see Eq.~(\ref{eq8})). The same conclusion holds for the diffusion
constant of the oxygen atoms, as can be seen from the open circles
in Fig.~\ref{fig8}.  Similar results have also been obtained for the
relaxation time of $F_s(q,t)$ for silicon as well as the various coherent
intermediate scattering functions~\cite{horbach_diss}.  It should be
noted, however, that this power law can be observed only in a
temperature range in which $\tau$ changes by about one decade. This
is significantly less than the range that has been observed for
simple liquids~\cite{lj_kob,roux89,vanmegen}, water~\cite{water},
molecular glassformers~\cite{molecular,kammerer}, or polymeric
systems~\cite{polymers}.  Thus we conclude that for this system the
corrections to the idealized version of the theory due to the hopping
processes make the observation of the power law rather difficult.

As mentioned above, the value of the exponent $\gamma$ should be
independent of the observable. In the inset of Fig.~\ref{fig8} we plot
the value of $\gamma$ for the three different wave-vectors as well as
the diffusion constant (point at $q=0$). We see that whereas the value
of $\gamma$ for intermediate values of $q$ is around 2.4, the one for the
diffusion constant is significantly smaller, $\gamma=2.05$. Thus again we
find that this quantity behaves somewhat anomalous with respect to the
ones at large $q$, and the reason for this is likely the one discussed
above. Also included in the inset, horizontal dashed line, is the value
of $\gamma$ as determined from a completely different approach, namely
the analysis of the $\beta$-relaxation regime (discussed below). We see
that this estimate of $\gamma$ is in very good agreement with the values
from the $\alpha$-relaxation times, thus supporting the prediction of
the theory that the exponent is not just a fit parameter but a quantity
with a fundamental theoretical meaning.

\begin{figure}[h]
\psfig{file=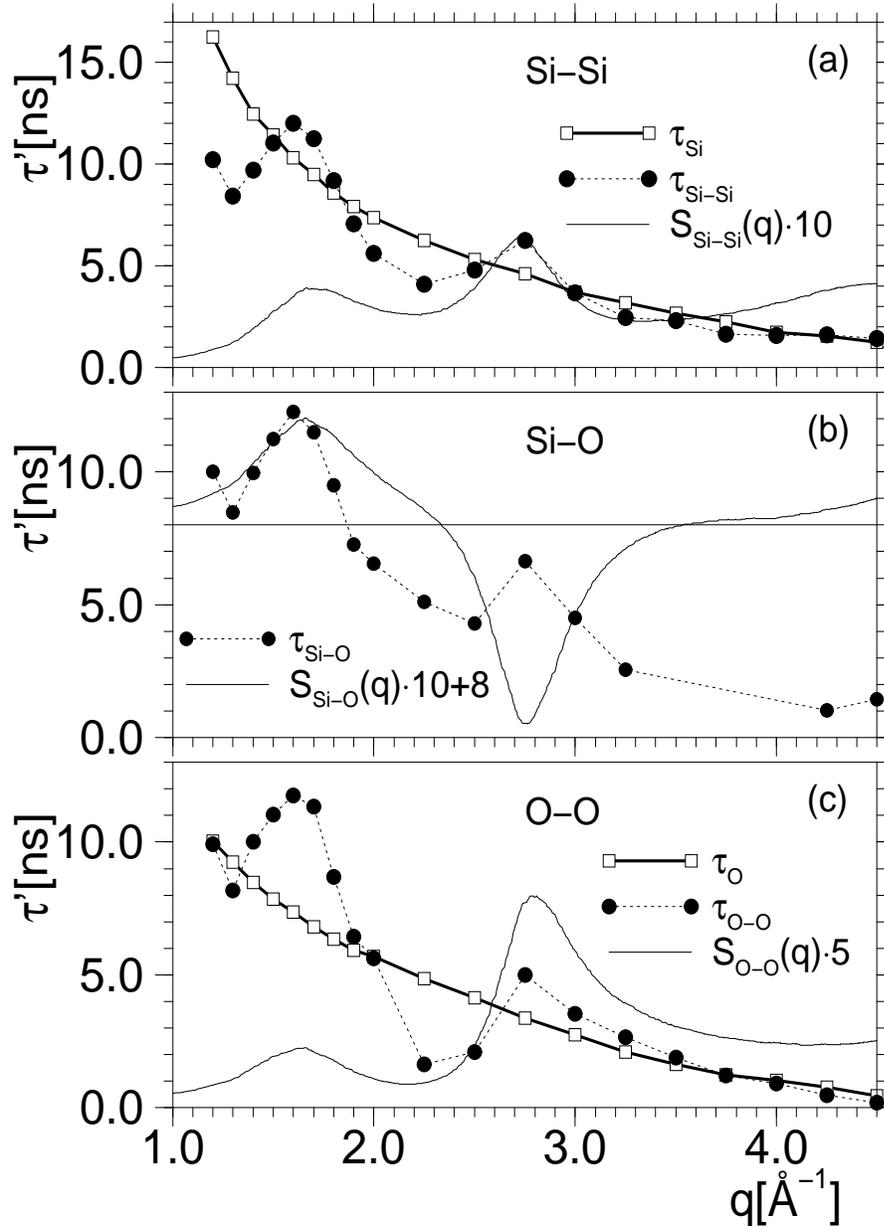,width=12cm,height=16cm}
\vspace{2mm}

\caption{
Wave-vector dependence of the $\alpha$-relaxation time $\tau'$ at $T=2750$~K
as determined from the incoherent and coherent intermediate scattering
functions (curves with symbols). Also included are the partial structure
factors from Ref.~\protect\cite{horbach99} (curves with no symbols).
}
\label{fig9}
\end{figure}

\begin{figure}[h]
\psfig{file=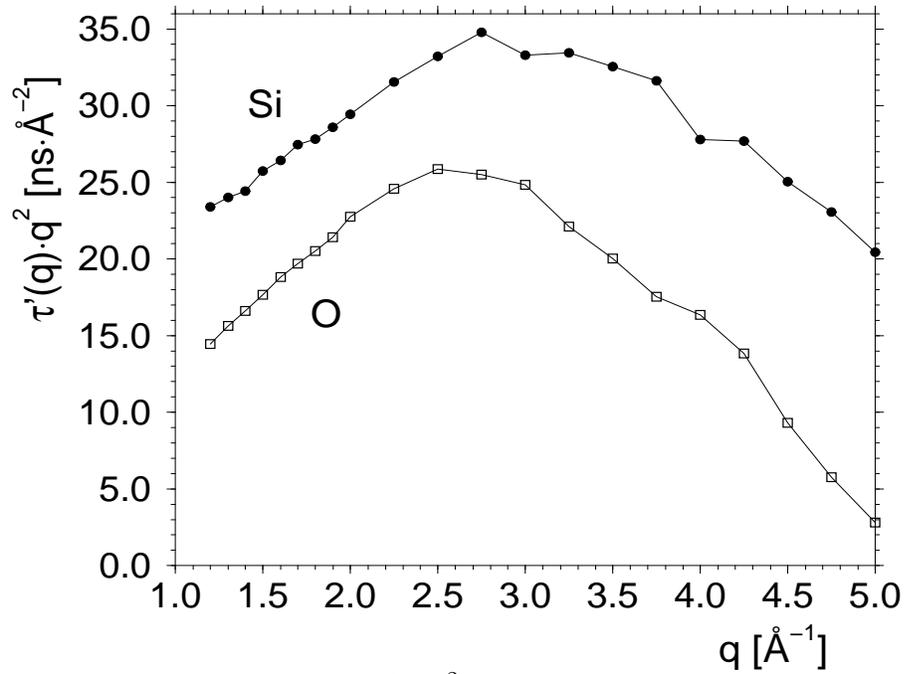,width=12cm,height=9cm}
\vspace{1mm}

\caption{
The product $\tau'(q) q^2$ as determined from the incoherent intermediate 
scattering functions as a function of $q$ at $T=2750$~K.
}
\label{fig10}
\end{figure}

We now discuss the wave-vector dependence of the $\alpha$-relaxation
times in more detail. Since we have seen, see Fig.~\ref{fig7}, that in
the whole temperature range investigated the temperature dependence of
$\tau$ is basically independent of $q$, it is sufficient to consider the
$q$-dependence at only one temperature. This is done in Fig.~\ref{fig9}
where we show the $q$-dependence of $\tau'$ for the two incoherent
as well as the three coherent functions. (Note, that for the reasons
discussed in the context of Fig.~\ref{fig6}, we show the $q$-dependence
of $\tau'$ instead of the one of $\tau$.) From the figure we see that
the relaxation time for the incoherent functions, panel (a) and (c),
decrease with increasing $q$. For a diffusive process, and hence also
for small $q$, this decrease is given by $1/q^2$~\cite{hansenmcdonald86}.
However, for the $q$-range considered here, i.e.~$q\ge 1.1$~\AA$^{-1}$,
one would expect that the behavior of $\tau$ is affected by the local
structure. That this indeed the case is shown in Fig.~\ref{fig10} where we
plot $\tau' q^2$ versus $q$.  The product $\tau' q^2$ can be interpreted
as an inverse $q$-dependent diffusion constant. Thus, generally speaking,
for $q$--values which correspond to length scales of nearest neighbor
distances, $\tau' q^2$ should be relatively large because then diffusion
processes should be relatively slow due to the local order.  As we see in
Fig.~\ref{fig10} $\tau' q^2$ exhibits an maximum around 2.7~\AA$^{-1}$ for
silicon as well as for oxygen which is close to the position of the main
peak in the static structure factor corresponding to the length scale of
a Si--O bond. It is of course remarkable that no such feature is found at
the position of the first sharp diffraction peak, at $q=1.7$~\AA$^{-1}$,
although such a feature is probably buried under the broad peak around
$q=2.7$~\AA$^{-1}$ (see Fig.~\ref{fig10}).  We can rationalize this
$q$-dependence from our previous observation~\cite{horbach99} that the
elementary diffusion step for the oxygen atoms is due to the breaking of
a Si--O bond (and a slightly more complicated process for the silicon
atoms) which shows that this length scale is of particular importance
for the dynamics. We note that a similar behavior is found also in the
MCT calculation for a hard sphere system~\cite{latz01}. In this
case one obtains a peak in $\tau q^2$ at a position which is close to
the first peak in the static structure factor.

In contrast to the monotonic decrease of the relaxation time for
the incoherent functions, the one for the coherent functions
shows an oscillatory behavior around the incoherent one, see
Fig.~\ref{fig9}. These oscillations are in phase with the partial
static structure factors, which have been included in the figures
as well. Whenever the modulus of $S_{\alpha\beta}(q)$ is large, the
corresponding $\tau(q)$ is large also, a phenomenon that is very similar
to the so--called de Gennes narrowing~\cite{hansenmcdonald86}.  Note,
however, that the de Gennes narrowing is a phenomenon observed at {\it
short} times and thus cannot be used to explain the $\tau(q)$ dependence
on the time scale of the $\alpha$-relaxation.  Finally we mention that
this type of dependence has also been found in a MCT calculation for a
hard sphere system~\cite{fuchs92} and recently in a neutron scattering
experiment of Toluene and m--Toluidine~\cite{alba_cm}.

\begin{figure}[h]
\psfig{file=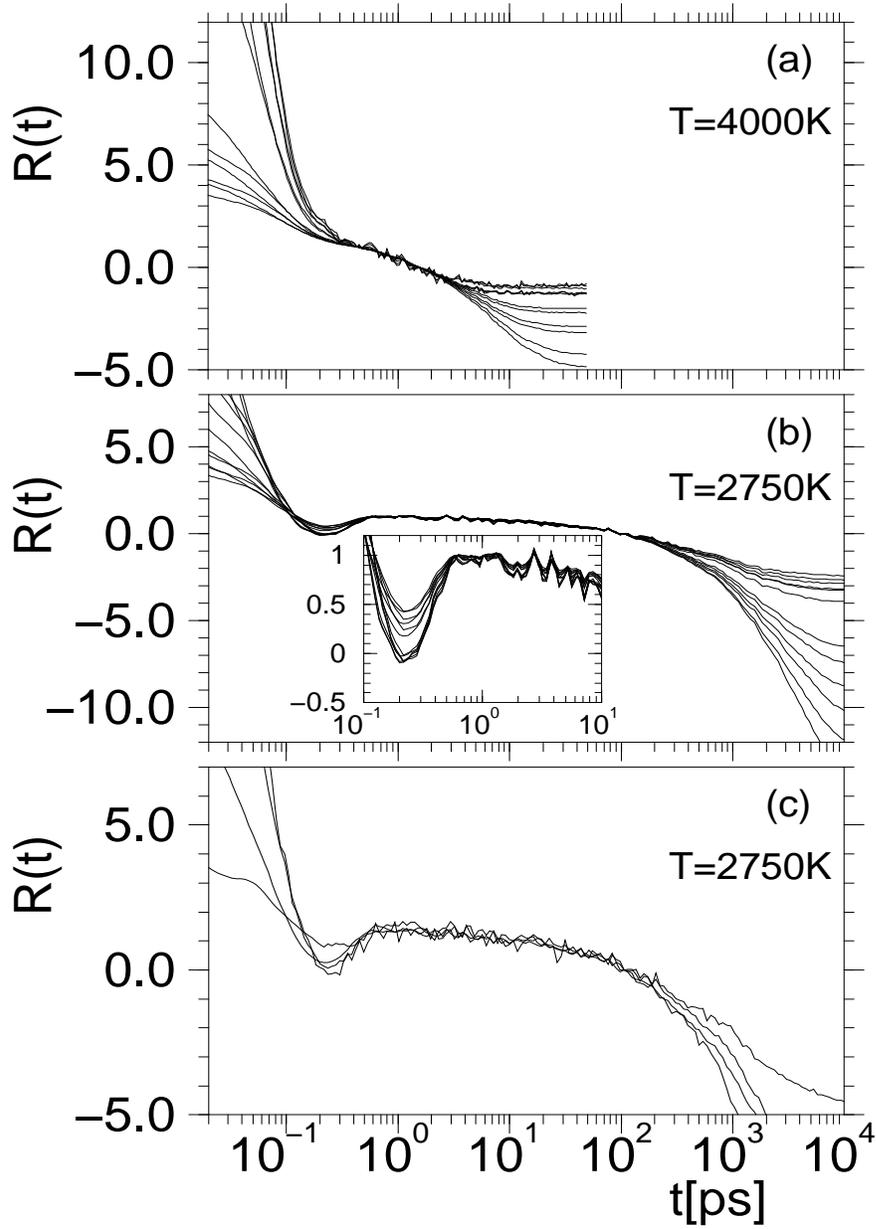,width=12cm,height=16cm}
\vspace{2mm}

\caption{
Time-dependence of the ratio $R(t)$, as defined in Eq.~(\protect\ref{eq10}),
for various correlators. a) $F_s(q,t)$ for silicon and oxygen at $q=1.7$,
2.2, 2.8, 4.43, 5.02 and 5.31~\AA$^{-1}$.  $T=4000$~K; b) the same set of
correlators as in a) but now at $T=2750$~K; c) four different correlators
(see main text for description) at $T=2750$~K.
}
\label{fig11}
\end{figure}

\begin{figure}[h]
\psfig{file=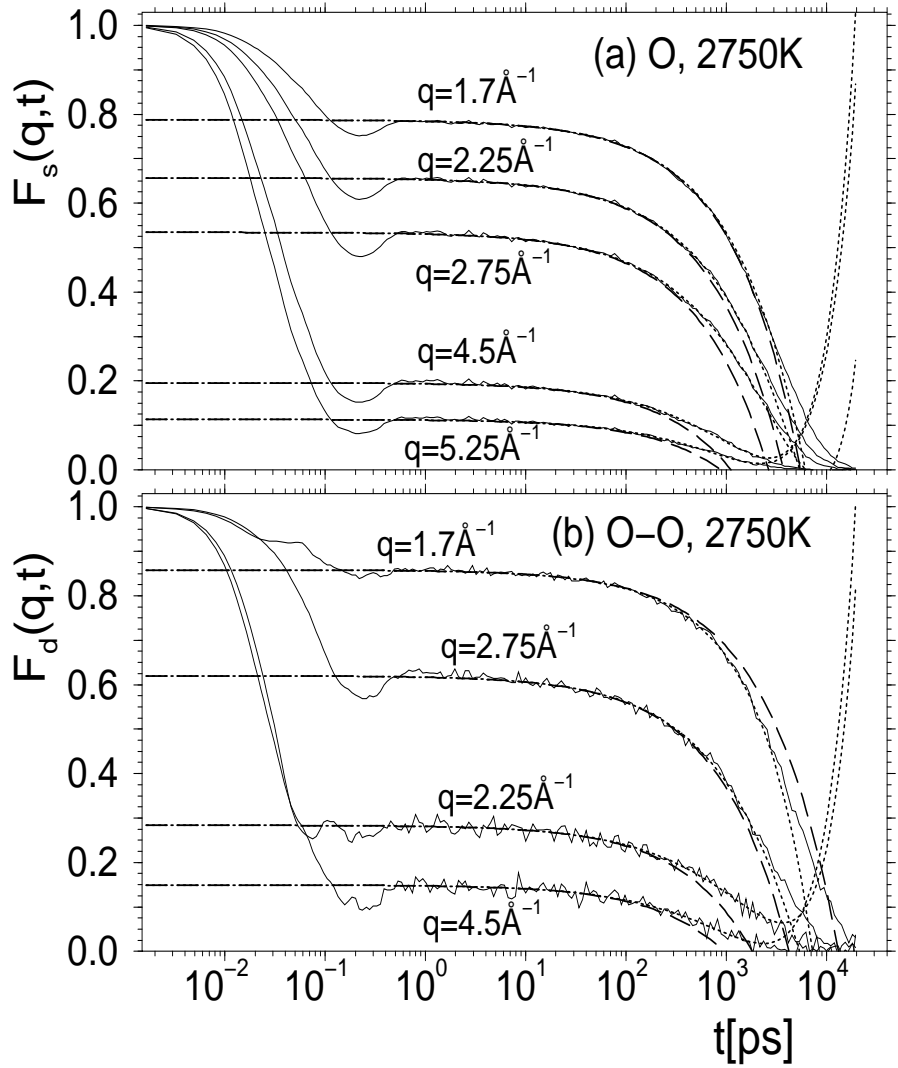,width=12cm,height=14cm}
\vspace{2mm}

\caption{
Time-dependence of $F_s(q,t)$ for oxygen, panel a), and $F_d(q,t)$
for the oxygen-oxygen correlation, panel b), for various wave-vectors (solid
lines). The dotted lines are fits the functional form given by
Eq.~(\protect\ref{eq13}). The dashed lines are fits with the functional form
given by Eq.~(\protect\ref{eq13}) without the last term.
}
\label{fig12}
\end{figure}

Having discussed so far mainly the $\alpha$-relaxation, we now turn
our attention to the $\beta$-relaxation, i.e.~the time window in which
the correlators are close to the plateau. MCT makes several interesting
predictions about this relaxation regime and therefore we will check in
the following to what extent they are valid for the present system. The 
first prediction of the theory is the so-called ``factorization property'',
which says that close to the plateau every time correlation function
$\phi_A(t)$ of an observable $A$ is given by

\begin{equation}
\phi_A(t)=\phi_A^c+h_{A} G(t),
\label{eq9}
\end{equation}

\noindent
where $\phi_A^c$ is the height of the plateau, $h_{A}$ is a constant,
and the whole time and temperature dependence is given by the function
$G(t)$, {\it which does not depend on the observable of $A$}. 
If the factorization property holds one can see immediately that the ratio

\begin{equation}
R_A = \frac{\phi_A(t)-\phi_A(t')}{\phi_A(t'')-\phi_A(t')}
\label{eq10}
\end{equation}

\noindent
is independent of $A$ if the times $t$, $t'$, and $t''$ are in the
$\beta$-relaxation regime, i.e.~where Eq.~(\ref{eq9}) holds. Note that
this prediction of the theory does not only hold for the idealized version
of the theory but also for the extended one, i.e.~if the hopping processes
mentioned in the Introduction are taken into account. Therefore it is
reasonable to check the validity of Eq.~(\ref{eq9}) not just above $T_c$,
but also below. This is done in Fig.~\ref{fig11} where we plot the ratio
$R(t)$ for a temperature above $T_c$, 4000K in panel (a), and below $T_c$,
2750K in panel (b). The correlation functions shown are $F_s(q,T)$ for silicon
and oxygen at $q=1.7$, 2.2, 2.8, 4.43, 5.02 and 5.31~\AA$^{-1}$. The times
$t''$ and $t'$ are 0.4~ps and 1.6~ps for $T=4000$~K and 11~ps and 106~ps for
$T=2750$K, respectively.

\mbox{From} Fig.~\ref{fig11}a we see that all the various curves
collapse nicely onto a master curve at intermediate times, i.e.~in the
$\beta$-relaxation regime. That this collapse is by no means trivial
is seen at short and long times since there we find no collapse at all,
thus showing that in general the shape of the correlation function does
depend on the observable. Only in the $\beta$-regime the correlators
show a system universal time dependence, in agreement with the theory.

The theory also predicts that with decreasing temperature the time
window over which the $\beta$-relaxation is observed should expand. That
this is indeed the case is demonstrated in Fig.~\ref{fig11}b, where
we show the same correlation functions as in panel (a), but now at
a lower temperature. From the figure we see that now the range over
which the correlators fall onto a master function has indeed increased
considerably, in agreement with the theory. Also included in this panel
is an enlargement of $R(t)$ at times where the correlator shows a dip,
due to the boson peak mentioned above. From this inset we see that on this
time scale the factorization property no longer holds since there the
correlation functions are dominated by the vibrational dynamics for which
Eq.~(\ref{eq9}) is not valid. 

So far we have considered only the incoherent intermediate scattering
functions. That the factorization property holds also for other time
correlation functions is shown in Fig.~\ref{fig11}c where we plot, for
$T=2750$K, the time dependence of $R(t)$ for $F_s(q,t)$ for oxygen at
$q=2.8$~\AA$^{-1}$, for $F_d(q,t)$ for Si-Si at $q=1.7$~\AA$^{-1}$ and
the functions $a_0^{\rm Si}(t)$ and $a_0^{\rm SiO}(t)$ which have been
proposed by Roux {\it et al.}~\cite{roux89} and which are defined as

\begin{equation}
a_0^{\rm Si}(t)=\int_0^{r_0^{\rm Si}} 4\pi r^2 G_s^{\rm Si}(r,t) dr
\label{eq11}
\end{equation}

\noindent
and 

\begin{equation}
a_0^{\rm SiO}(t)=\int_0^{r_0^{\rm SiO}} 4\pi r^2 \left[ G_d^{\rm
SiO}(r,t)-1\right] \left[ G_d^{\rm SiO}(r,0)-1 \right] dr .
\label{eq12}
\end{equation}

\noindent
Here $G_s^{\rm Si}(r,t)$ and $G_d^{\rm SiO}(r,t)$
are the self and distinct part of the van Hove correlation
functions~\cite{hansenmcdonald86} which have been calculated and discussed
in Ref.~\cite{horbach99}. The values of $r_0^{\rm Si}$ and $r_0^{\rm
SiO}$ are 0.4~\AA~and 2.35~\AA, respectively. Thus we see that the four
time correlation functions considered are of very different nature
and their overall relaxation behavior is certainly not the same. From
Fig.~\ref{fig11}c we see, however, that in the $\beta$-relaxation regime
the four functions fall nicely onto a master curve. Therefore we find
that also for these correlators Eq.~(\ref{eq9}) holds, which is hence
strong evidence that for the present system the factorization property
is a general property of a vary large class of time correlation functions.

Having shown now that in the $\beta$-relaxation regime the shape of a
time correlation function $\phi_A(t)$ is independent of the observable $A$
considered, we continue now to study this shape in more detail. For
this we make use of a further prediction of MCT, namely that the time
dependence of $\phi_A(t)$ at long times is not arbitrary, but given by a
sum of power laws of the form:

\begin{equation}
\phi_A(t)= \phi_A^c-h_A t^b+ h_A^{(2)}t^{2b}+ \cdots .
\label{eq13}
\end{equation}

\noindent
Here $\phi_A^c$ is the height of the plateau, often also called the
non-ergodicity parameter, and $h_A$ (also called critical amplitude)
and $h_A^{(2)}$ are constants. According to MCT the value of the
exponent $b$, also called von Schweidler exponent, is {\it independent}
of $A$, i.e.~it is a system-universal quantity. The first two terms
on the right hand side are the so-called von Schweidler law. Note that
Eq.~(\ref{eq13}) is an expansion of the correlator in terms of $t^b$,
i.e.~it is assumed that the third term is smaller than the second one.
However, it has been suggested by the theory~\cite{mct_corrections} that
in order to determine reliably the coefficient $h_A$ it is usually
necessary to take into account these correction as well.  In order to
check whether the Ansatz~(\ref{eq13}) gives a good description of our
correlation functions we have fitted them in the $\beta$-relaxation
regime with this functional form. The temperature we used was 2750~K,
since this gives us the longest plateau and hence allows us to determine
the fit parameters with the highest precision. The result of these fits
are shown in Fig.~\ref{fig12} for the case of the incoherent function
for the oxygen atoms, panel (a), and the coherent function for the O-O
correlation, panel (b). For each correlator (solid lines) two fits are
shown: The one in which all three terms in Eq.~(\ref{eq13}) were used
(dotted lines) and the one in which the third term was neglected (dashed
line). The difference between these two fits are seen at long times where
the fits with the correction terms are able to describe the correlator
for about one decade more than if these terms are not taken into account.
That the quality of the fits improves significantly by considering the
correction term $\propto t^{2b}$ was first shown by Sciortino {\it et al.}~for
the case of water~\cite{water}.
We emphazise that for these fits the value of $b$ was kept fixed for all
the different correlators, i.e.~it was a global fit parameter. The value
we obtained for $b$ was $0.62 \pm 0.02$. It can be seen from the figure
that, if the correction terms are taken into account, the resulting
fits are very good and describe the data over more than three decades
in time. Therefore we conclude that this prediction of the theory,
i.e.~Eq.~(\ref{eq13}), holds for the present system.

A further interesting prediction of MCT is that there exists a one-to-one
correspondence between the value of the von Schweidler exponent $b$ from
Eq.~(\ref{eq13}) and the exponent $\gamma$ of the $\alpha$-relaxation
time in Eq.~(\ref{eq8})~\cite{mct,gotze99}. According to the theory one
can start with $b$ and determine from

\begin{equation}
\frac{\Gamma^2(1+b)}{\Gamma(1+2b)} = \frac{\Gamma^2(1-a)}{\Gamma(1-2a)}
\label{eq14}
\end{equation}

\noindent
the value of the parameter $a$ and then use the relation

\begin{equation}
\gamma=\frac{1}{2a}+\frac{1}{2b}
\label{eq15}
\end{equation}

\noindent
to calculate $\gamma$. (Here $\Gamma(x)$ is the $\Gamma$-function.) If
this is done for our value $b=0.62$, one obtains $\gamma=2.35$. Since
we have determined the value of $\gamma$, as defined by Eq.~(\ref{eq8})
also directly from the temperature dependence of the $\alpha$-relaxation
time, we can compare it now to this theoretical value.  This is done
in the inset of Fig.~\ref{fig8} and, as discussed already above, the
theoretical value agrees well with the ones from $\tau(T)$ for finite
values of $q$. Thus we conclude that also this highly non-trivial
prediction of the theory is satisfied quite accurately.

\begin{figure}[h]
\psfig{file=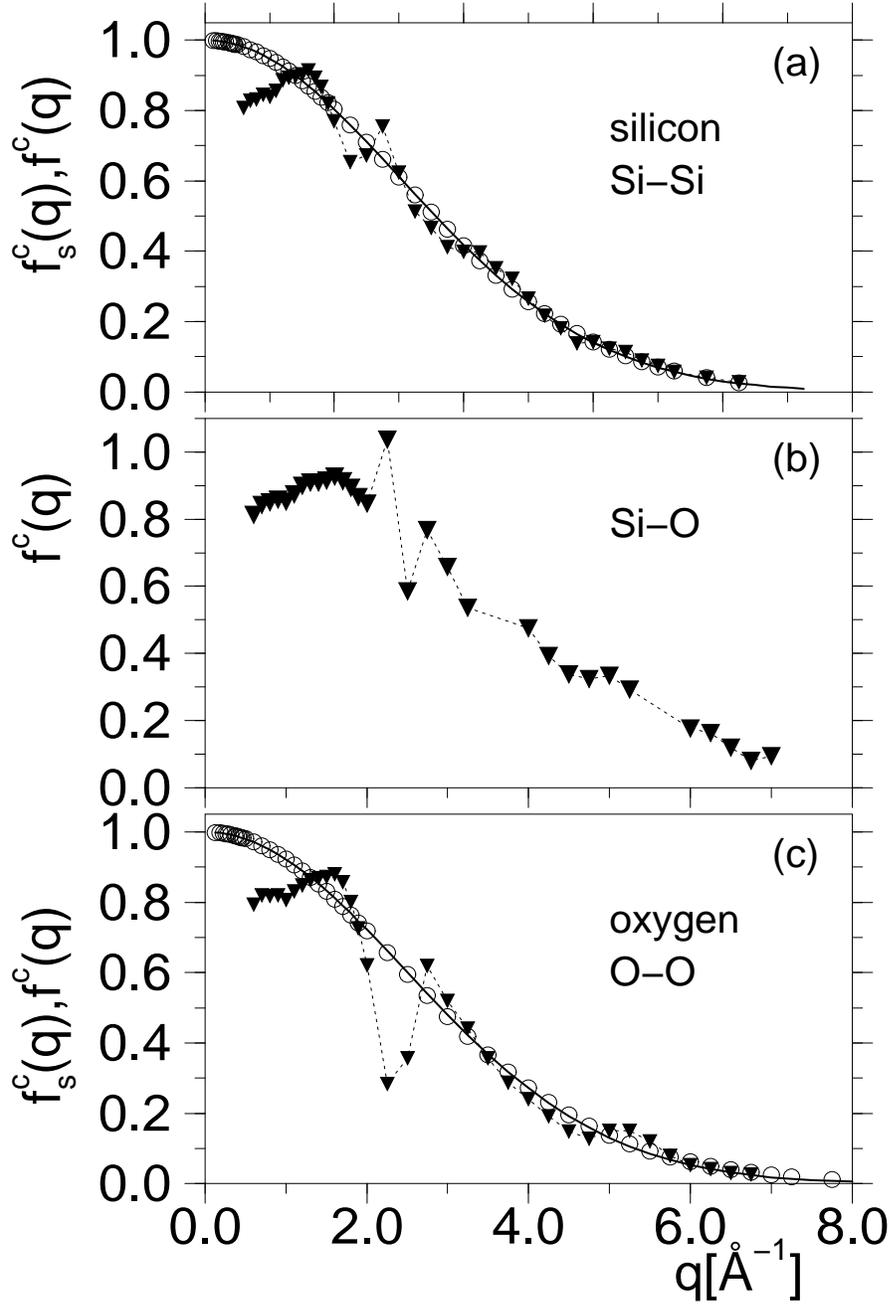,width=12cm,height=17cm}
\vspace{2mm}

\caption{
Wave-vector dependence of the nonergodicity parameters. a) $f_s^c(q)$
from $F_s(q,t)$ for silicon and $f^c(q)$ from $F_d(q,t)$ for the
silicon-silicon correlation; b) $f^c(q)$ from $F_d(q,t)$ for the
silicon-oxygen correlation; c) $f_s^c(q)$ from $F_s(q,t)$ for oxygen and
$f^c(q)$ from $F_d(q,t)$ for the oxygen-oxygen correlation.  The solid
lines in a) and b) are a fit with a Gaussian to the incoherent data.
}
\label{fig13}
\end{figure}

In the last part of this section we discuss the wave-vector dependence
of the parameters that where obtained from the fits of the correlators in
the $\beta$-regime to the functional form given by Eq.~(\ref{eq13}). For
this we have made fits to the two incoherent and the three coherent
intermediate scattering functions at $T=2750$K for a wide range of
wave-vectors. The dependence of the non-ergodicity parameters is shown in
Fig.~\ref{fig13}. For the case of the incoherent function, $f_s^c(q)$, we
see that the $q$-dependence is Gaussian like. A fit with the functional
form $\exp(-q^2 r_s^2)$ gives a very good description of the data
(see solid lines in panels (a) and (c)). For the localization length
$r_s$, i.e.~the size of the cage, we obtain for silicon and oxygen the
values 0.23~\AA~and 0.29~\AA, respectively.  This has to be compared with
the typical nearest neighbor distance between silicon and oxygen, which
is $d=1.6$\AA~\cite{horbach99,mozzi69}. Thus we find that $r_s$
is around $0.14d$ and $0.18d$ for silicon and oxygen, respectively. This
size is somewhat larger than the one found in most solids which, in
agreement with the Lindemann criterion~\cite{lindemann10}, have usually
$r_s \approx 0.1d-0.14d$. The reason for this difference might lie in the fact
that SiO$_2$ is a very open network and thus the atoms can vibrate with
a relatively large amplitude without approaching each other too much.

The wave-vector dependence of the non-ergodicity parameter for
the coherent functions shows oscillations which are in phase
with the corresponding partial structure factor. For the case of
Si-Si and O-O they oscillate around the corresponding $f_s^c$ and
become basically indistinguishable from them at large $q$. These
results are in qualitative agreement with the ones obtained for other
systems~\cite{water,lj_kob,molecular,kammerer,polymers} 
which shows that from this point
of view the present open network system does not behave differently as a
simple liquid or a polymer. As already mentioned in the Introduction, it
has very recently been shown that the $q$-dependence of the non-ergodicity
parameters $f^c(q)$ shown here can be calculated also reliably with
the theory {\it without any fit parameter}. Since the agreement was also good
at large wave-vectors, and since we see now that the coherent function
oscillates around the incoherent one, we expect that the theory is indeed
able to give a reliable estimate of the size of the cage also for this
network forming system.

\begin{figure}[h]
\psfig{file=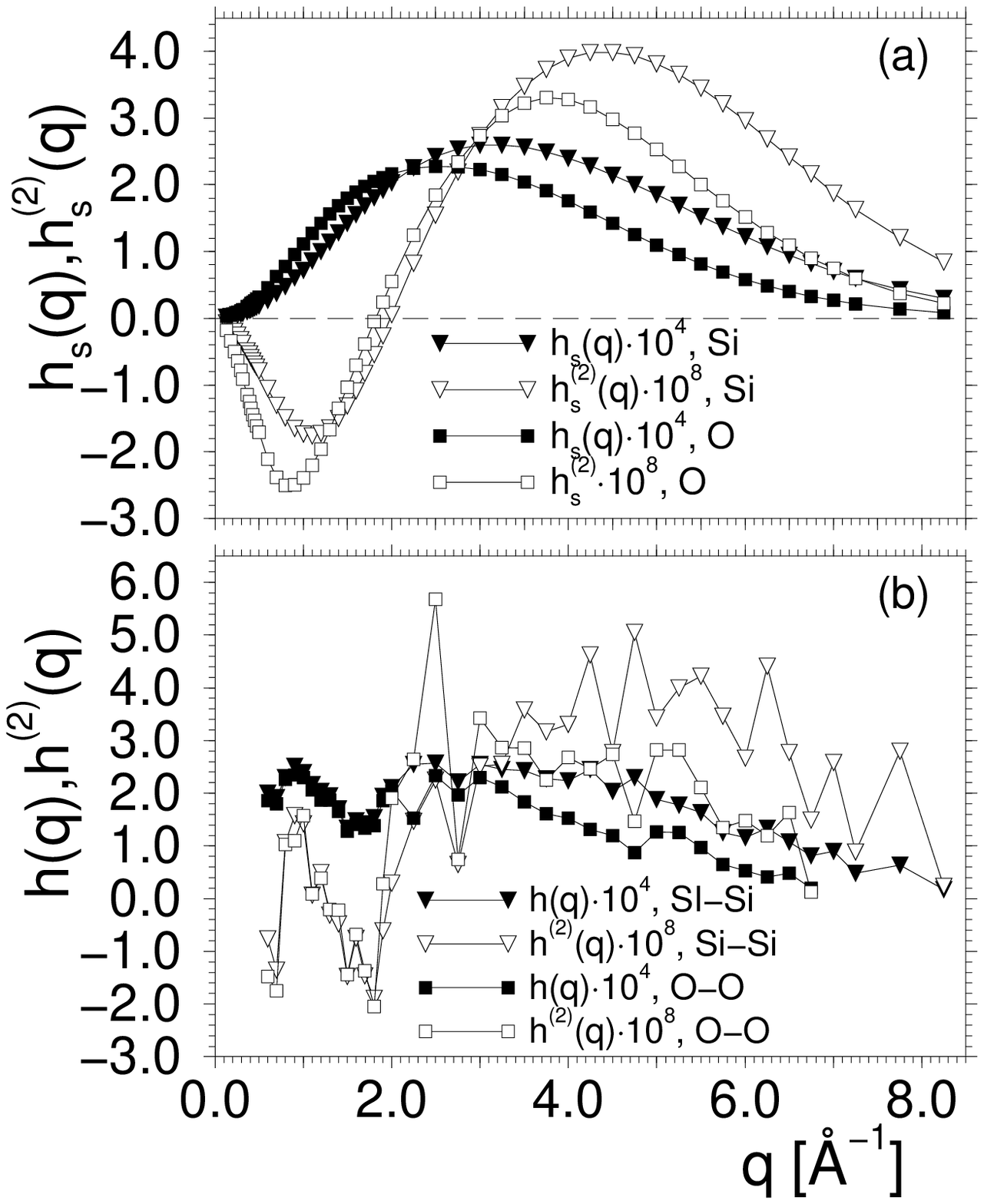,width=12cm,height=15cm}
\vspace{2mm}

\caption{
Wave-vector dependence of the critical amplitude $h(q)$ and the first
correction term $h^{(2)}(q)$. a) Incoherent functions for silicon and oxygen;
b) Coherent functions for the silicon-silicon and the oxygen-oxygen
correlations.
}
\label{fig14}
\end{figure}

The wave-vector dependence of the remaining parameters are shown in
Fig.~\ref{fig14}. In panel (a) we show the critical amplitude $h_s$ and
the first correction $h_s^{(2)}$ for the two incoherent functions. From
this figure we recognize that from a qualitative point of view the curves
are very similar to the ones calculated within the MCT for a system of
hard spheres~\cite{fuchs92}. In particular we find that $h_s^{(2)}$
shows at small wave-vectors a negative minimum and subsequently a
positive maximum at significantly larger $q$. In between it becomes zero at
$q=1.98$~\AA$^{-1}$ and 1.82~\AA$^{-1}$ for the case of silicon and oxygen,
respectively. This is close to the first peak in the structure factor,
in agreement with the results for the hard sphere system~\cite{fuchs92}.
This results shows that also from this point of view the theoretical
calculation is qualitatively correct. Note that the fact that
$h_s^{(2)}(q)=0$ means that for this wave-vector the von Schweidler law,
i.e. the second term in Eq.~(\ref{eq13}) can be seen over the broadest
possible time-window, since the first correction to this law is zero.

Also the $q$-dependence of $h(q)$ and $h^{(2)}(q)$ for the collective quantities, 
Fig.~\ref{fig14}b,
is in qualitative agreement with the MCT results for a hard sphere
system~\cite{fuchs92}. E.g. $h^{(2)}(q)$ shows at the location of the
first peak a minimum and its value is negative. Thus we see that the
theory is able to rationalize the behavior of this function, at least
in a qualitative way.  Of course it would be a more stringent test to
do the MCT calculation directly for the silica system instead for hard
spheres. However, this type of calculation is currently still very
demanding and thus has to be left as a future project.

\section{Summary and discussion}
\label{sec4}
In this paper we have presented the results of molecular dynamics computer
simulations of a model for silica, the archetype of a strong glass former.
The goal of this study was twofold: On the one hand to investigate in
detail the relaxation dynamics of this system in the temperature regime
where this dynamics is slow. On the other hand we wanted to check to
what extent the mode-coupling theory of the glass transition is able to
describe this dynamics at low temperatures. In these investigations we
mainly focussed on the time-and temperature dependence of the various
intermediate scattering functions for different wave-vectors. We found that
these functions show nicely the slowing down of the system upon cooling in
that they change their shape from a single exponential at high temperatures
to a two-step relaxation at low temperatures. We emphasize that this two-step
relaxation is already seen at temperatures which are about 30\% above the
melting temperature of the system, which shows that in order to show a slow
dynamics the system does not have to be supercooled. 

In a previous paper it was demonstrated that this system
shows a crossover of the dynamics from a flow-like motion at
intermediate and high temperatures to a hopping-like motion at low
temperatures~\cite{horbach99}. The temperature of this crossover
can be identified with the critical temperature of MCT and is around
3330~K. We now find that the time-temperature superposition principle
holds for temperatures around $T_c$ and that time correlation functions
are stretched. For significantly higher and lower temperatures, the TTSP
is violated in that the correlators become more exponential-like. This
observation is also in accord with the existence of the above mentioned
crossover from a flow-like dynamics (i.e.~collective motion which leads
to a stretched exponential relaxation) to a dynamics in which hopping
dominates (i.e.~single particle dynamics which leads to an exponential
relaxation). These results are in qualitative agreement with the
prediction of the theory~\cite{gotze87}.

The temperature dependence of the $\alpha$-relaxation times are
qualitatively similar to the one of the diffusion constant in that also
they show a crossover from an Arrhenius law at low temperatures to a
weaker temperature dependence at higher $T$. In this latter temperature
range the $T$-dependence of $\tau$ can be fitted well with a power law
with the same critical temperature as found in Ref.~\cite{horbach99}.
However, the exponent $\gamma$ of this power law that we find for the
intermediate scattering function is significantly higher than the one
for the diffusion constant, in contradiction with the leading order prediction of
the theory. Similar to the findings in simple liquids, the reason for
this discrepancy might be the existence of dynamical heterogeneities,
i.e.~due to the fact that the temperature dependence of the diffusion
constant is weaker than expected.

According to the theory the time-temperature superposition principle
should hold also in the $\beta$-relaxation regime. We find that a
plot of the correlation functions versus rescaled time does not
lead to a collapse of the curves onto a master function in this
time window. The reason for this is likely due to the presence of a
very pronounced vibrational dynamics, related to the boson peak, that
extends to relatively low frequency, i.e.~influences the dynamics also
at relatively long times. It has been shown before that the presence of
a pronounced microscopic dynamics may disturb the $\beta$-relaxation
dynamics predicted by MCT so strongly that the asymptotic results
cannot be observed anymore~\cite{franosch98}. Therefore it is not
surprising that for the present system the TTSP does not hold in the
$\beta$-regime. (Note, that the reason that the TTSP is nevertheless
observed in the $\alpha$-regime is likely the fact that the latter regime
is much less affected by the microscopic vibrations than the former
since on the time scale of the $\alpha$-relaxation the vibrations have
finally been damped out.) However, by calculating the function $R(t)$
from Eq.~(\ref{eq10}) we can show that the time dependence of a wide class
of correlation functions is the same for times in the $\beta$-relaxation
regime. In agreement with the theory this dependence is a sum of fractal
power laws. We also find that the value of the von Schweidler exponent
$b$ in these laws obeys the connection proposed by MCT between $b$ and
the exponent $\gamma$ of the $\alpha$-relaxation time, if one uses the
$\gamma$ as determined from $\tau(T)$ from the intermediate scattering
function at {\it finite} wave-vectors. This result shows that neither
$\gamma$ nor $b$ are mere fit parameters but instead that they have a
more fundamental theoretical meaning.

By fitting the coherent and incoherent intermediate scattering
functions in the $\beta$-relaxation regime with the functional form
proposed by MCT we have determined the wave-vector dependence of
the various nonergodicity parameters, the critical amplitudes,
and the correction terms. From a qualitative point of view these
dependencies look very similar to the ones found in simple liquids or
simple molecules~\cite{water,lj_kob,molecular,kammerer}.  Since the latter
systems are all fragile glass formers we thus conclude that for these
quantities there is no qualitative difference between fragile and strong
glass formers. Furthermore it has to be mentioned that these dependencies
are also in qualitative agreement with the ones predicted by MCT {\it for
the simpler systems}, thus giving evidence that the theory is indeed able
to describe them correctly. 

In summary, we can say that MCT allows to understand many aspects of the
dynamics of this strong glass former. Moreover, in Ref.~\cite{sciortino01}
it was shown that the $q$-dependence of the Debye--Waller factor is 
predicted well by MCT. To what extent this
is true also for the other observables studied in the present paper is
not yet known and it is certainly important to check this in the future.
It is of interest that we find that some of the predictions of the theory,
such as the factorization property, hold also below $T_c$, i.e.~in a
temperature regime where the hopping processes are important. For the
present system we find, however, that in the $\beta$-relaxation regime
the time-temperature superposition principle does not hold. Whether this
is a real failure of the theory for strong glass formers or whether this
is just a side effect of the strong vibrational dynamics at short times,
can unfortunately not be decided. Hence it would be of great interest
to investigate this question for a strong glass former which does not
show such a pronounced vibrational dynamics.

Acknowledgments: We thank K. Binder, W. G\"otze, M. Fuchs, A. Latz, and 
F. Sciortino for useful discussions. This work was supported by the BMBF 
Project 03~N~8008~C and the DFG under SFB~262. Part of this work was done at 
the RUS in Stuttgart.

\begin{references}
\bibitem{footnote1}
The reason for choosing 0.1 instead of the value discussed earlier,
$1/e$, is that for large wave-vectors the height of the plateau is
smaller than $1/e$. Thus with the old (=usual) definition one would
measure a microscopic time instead of the $\alpha$-relaxation time. As
long as the TTSP is valid, the precise nature of the definition affects
the value of the $\alpha$-relaxation time only via a prefactor.
\end{references}
\end{document}